\documentclass[runningheads,natbib]{svjour2}
\bibpunct{[}{]}{;}{n}{}{,} % to get "[numbered]" references from natbib
\smartqed  % flush right qed marks, e.g. at end of proof
\usepackage{graphicx}

\newcommand{\hi} {{\rm H}\,{\small\rm I}}
\newcommand{\kms} {\,{\rm km\,s}^{-1}}
\newcommand{\kmskpc} {\,{\rm km\,s}^{-1}\,{\rm kpc}^{-1}}

\newcommand{\mo}{\,{M}_\odot}
\newcommand{\moyr}{\,{M_\odot\,\rm yr}^{-1}}
\newcommand{\cm}{\,{\rm atoms}~{\rm cm}^{-2}}
\newcommand{\mopc}{M_\odot\,{\rm pc^{-2}}}
\newcommand{\gsim}{\lower.7ex\hbox{$\;\stackrel{\textstyle>}{\sim}\;$}}
\newcommand{\lsim}{\lower.7ex\hbox{$\;\stackrel{\textstyle<}{\sim}\;$}}

\journalname{Astronomy and Astrophysics Review}

\begin{document}

\title{Cold gas accretion in galaxies}


\author{Renzo Sancisi \and
  Filippo Fraternali \and
  Tom Oosterloo \and
  Thijs van der Hulst
}


\institute{R. Sancisi \at
              Osservatorio Astronomico di Bologna,
              Via Ranzani 1, I-40127 Bologna, Italy\\
              \email{sancisi@bo.astro.it} \\
              and Kapteyn Astronomical Institute,
              Postbus 800, NL-9700 AV Groningen, The Netherlands\\
              \email{sancisi@astro.rug.nl}
              \and
              F. Fraternali \at
              Astronomy Department, Bologna University,
              via Ranzani 1, I-40127 Bologna, Italy \\
              \email{filippo.fraternali@unibo.it}
              \and
              T. Oosterloo \at
              ASTRON,
	      Postbus 2, NL-7990 AA Dwingeloo, The Netherlands\\
	      and Kapteyn Astronomical Institute,
              Postbus 800, NL-9700 AV Groningen, The Netherlands\\
              \email{oosterloo@astron.nl} 
              \and             
              J.M. van der Hulst \at
              Kapteyn Astronomical Institute,
              Postbus 800, NL-9700 AV Groningen, The Netherlands\\
              \email{vdhulst@astro.rug.nl}.
}

\date{Received: }
% The correct date will be entered by the editor

\maketitle

\begin{abstract}

Evidence for the accretion of cold gas in galaxies has been rapidly 
accumulating in the past years. \hi\ observations of galaxies and 
their environment have brought to light new facts and phenomena 
which are evidence of ongoing or recent accretion:

1) A large number of galaxies are accompanied by gas-rich dwarfs
or are surrounded by \hi\ cloud complexes, tails and filaments.
This suggests ongoing minor mergers and recent arrival of external gas.
It may be regarded, therefore, as direct evidence of cold gas accretion 
in the local universe. It is probably the same kind of phenomenon 
of material infall as the stellar streams observed in the halos of our 
galaxy and  M\,31.

2) Considerable amounts of extra-planar \hi\ have been found in nearby 
spiral galaxies. While a large fraction of this gas is undoubtedly
produced by galactic fountains, it is likely that a part of it is of 
extragalactic origin. Also the Milky Way has extra-planar gas complexes: 
the Intermediate- and High-Velocity Clouds (IVCs and HVCs).

3) Spirals are known to have extended and warped outer layers of \hi.
It is not clear how these have formed, and how and for how long the warps 
can be sustained. Gas infall has been proposed as the origin.

4) The majority of galactic disks are lopsided in their morphology 
as well as in their kinematics. Also here recent accretion has been 
advocated as a possible cause.

In our view, accretion takes place both through the arrival and 
merging of gas-rich satellites and through gas infall from the 
intergalactic medium (IGM). The new gas could be added to the halo or be 
deposited in the outer parts of galaxies and form reservoirs for 
replenishing the inner parts and feeding star formation.
The infall may have observable effects on the disk such as bursts of 
star formation and lopsidedness.

We infer a mean ``visible'' accretion rate of cold gas in galaxies 
of at least $0.2 \moyr$. 
In order to reach the accretion rates needed to sustain the observed 
star formation ($\approx 1 \moyr$), additional infall of large amounts of gas 
from the IGM seems to be required.

\end{abstract}

\section{ Introduction}
\label{intro}

Gas accretion plays a fundamental role in the evolution of galaxies. 
Fresh supplies of gas are needed for the ongoing process 
of star formation. Such a process of galaxy ``nurture'' is 
expected to continue to the present day.
The importance and  role of gas infall for the evolution of disk galaxies 
have been recognized for many years \cite{lars72, lars80, tinsley80, tosi88}.

The rate of star-formation in the solar neighborhood has been remarkably
constant over the Milky Way's life \cite{twarog,BinneyDB}, which suggests that
the gas consumed by star formation has been replaced by accretion.
Steady accretion of metal-poor gas would also explain the discrepancy
between the observed stellar metallicity distribution in the solar
neighbourhood and that predicted by closed-box models of chemical
evolution \cite{tinsley81,matteucci}.
It is not clear how much new gas is needed to sustain star formation.
The star formation rate (SFR) varies throughout the Galactic disk and 
from galaxy to galaxy. Its value is still very uncertain. 
Here, we assume a reference global value of $1 \moyr$ both for 
the average SFR and for the required gas accretion rate.

Several arguments suggest that most of the baryons in the local 
universe still reside in the intergalactic medium 
\cite{whi91,fuku04,som06}. Out of this medium 
galaxies are expected to grow through a series of infall events ranging from 
a small number of major mergers down to an almost continuous
infall of dwarf galaxies and gas clouds, the latter being more and more 
important at low redshifts \cite{Bond,LaceyCole}.
Recent, high-resolution cosmological simulations show that there
are two modes of accretion: hot accretion, mainly around massive structures,
and cold accretion (clouds, streams or filaments) 
for galaxies with lower halo masses, which correspond to 
the population of star forming galaxies \cite{dekel06}.
These arguments also point to a rate of gas accretion for galaxies which is 
very close to their star-formation rate \cite{keres05}.

Direct observational evidence of accretion actually taking place has, 
however, been difficult to obtain. 
The study of \hi\ in the Milky Way and in external galaxies has played 
a central role. For several decades \hi\ observations have given valuable information on gaseous content, structure and kinematics of galaxies and on 
their environment. In particular, the \hi\ rotation curves have provided 
the crucial evidence for dark matter in spiral galaxies. 
New facts have been revealed, especially in recent deep observations, 
which now constitute the best evidence 
for cold gas accretion. We briefly review them here.

First (Section \ref{interactions}), we focus on those 
phenomena -- interactions, minor mergers, peculiar \hi\  
structures around galaxies -- which in our opinion point directly at ongoing 
or recent processes of accretion. We believe that the stellar merger remnants recently discovered in the halo of our galaxy (e.g.\ Sgr Dwarf) \cite{ibata94} and of M 31 \cite{ibata01,ferg02,mccon03}, and other faint optical features found  around some nearby galaxies \cite{mal97, shang98}, are manifestations of the same  phenomena as those revealed by \hi\ observations and described here.

Subsequently we draw attention to other aspects of the structure and 
kinematics of \hi\ in galaxies -- the extra-planar gas, the extended and warped 
outer layers and the lopsidedness --  which may be part of the accretion process.
The connection between accretion and the above phenomena, however, is not 
entirely clear and most of the evidence is indirect.  
The extra-planar gas (Section \ref{extraplanar}), although in part undoubtedly a product of galactic fountains \cite{fb06}, 
must also have a component which originated from infall or minor mergers
\cite{oos07,fb08}.
In our discussion of extra-planar \hi\ we include, 
together with the recent evidence from external galaxies, 
the long-known High-Velocity Clouds (HVCs),
most of which are now conclusively regarded as a Milky Way 
halo population and direct evidence for infall of 
intergalactic gas \cite{wak07,wak08}.  
The warped outer \hi\ layers (Section \ref{warps}) 
of spirals may also be the result of infall and form a source of continuous 
supply of fresh gas for the inner disks.
Finally (Section \ref{lopsidedness}), also the 
lopsided \hi\ morphology and kinematics and the asymmetric optical images 
of a large number of disks may have originated from recent minor mergers 
or large-scale cosmological gas accretion \cite{bou05}.  
The question of the intergalactic origin of the infalling gas is addressed in 
Section~\ref{IGM}.

\section {Interactions and minor mergers}
\label{interactions}

\subsection{Dwarf companions and peculiar structures}

There are several cases of multiple systems with similar mass (e.g.\ 
M\,81-M\,82-NGC\,3077 \cite{yun94},
NGC\,4631-4656-4627 \cite{rand94},
NGC\,5194(M51)-5195 \cite{rots90})
which show heavily disturbed \hi\ images 
with associated cloud complexes, long tails, bridges and ring-like 
structures.  For many of these systems,
it is the peculiar \hi\ picture that unmistakably
points at the ongoing strong tidal interactions and major mergers.

Here, however, we draw attention to galaxies which are interacting with gas-rich 
dwarf companions and to galaxies with no apparent interaction 
but with peculiar \hi\ structures and/or kinematics. All these systems can 
be considered as minor mergers at different stages. 
Those with companions show \hi\ tails and bridges indicating that an 
interaction is indeed taking place. The others, with no visible companions,  
have peculiar features in their \hi\ structure and kinematics, 
especially in their outer parts, which are reminiscent 
of interacting systems. They may, therefore, have had some recent 
encounter and may be in an advanced stage of merging 
\cite{san99a,san99b}. 
However, as far as we know, they may also be the result of the infall 
of intergalactic gas clouds. 

A number of representative cases are listed in Table \ref{t_interactions}.
This list is by no means complete. A compilation of a large number  
of \hi\ maps of peculiar galaxies, which includes many cases of the kind 
discussed here, is provided by ``An \hi\ Rogues Gallery'' \cite{hibba01}.

\begin{table}[t]
\caption{Galaxies with dwarf companions and/or peculiar \hi\ structures}
\begin{center}
\label{t_interactions}
\begin{tabular}{llcc}
\hline\noalign{\smallskip}
Object & Features & Masses  &  Reference \\[3pt]
       & &       ($10^8 \mo$)   &            \\[3pt]
\tableheadseprule\noalign{\smallskip}

IC10	 & - & - & \cite{hibba01}\\
NGC\,210 & tail & 6 & $^a$\\
NGC\,262 (Mkn\,348)& tail & $>20$ & \cite{heck82, sim87}\\
NGC\,628 & high-velocity complexes & 2 & \cite{kam92}\\
NGC\,925 & tail & 3 & $^b$\\
NGC\,1023 (Arp 135) & tail/ring & 10 &	\cite{san84}\\
NGC\,1961 (Arp 184) & wing & $\sim$54 & \cite{sho82}\\
NGC\,2146 & cloud & 46 & \cite{fish76, tara01} \\
NGC\,2782 (Arp 215) & plume & $\sim$10 & \cite{smi94} \\
NGC\,2985 (UGC\,5253)& tail, interaction & $>3.6$ & $^b$\\
NGC\,3067 & plume & 2 & \cite{car92}\\
NGC\,3310 (Arp 217) & tails & 5 & \cite{mul95, kre01}\\
NGC\,3359 & companion/bridge & $9.6$ & \cite{kam94, hulst05}\\
NGC\,4027 (Arp 22) & companion/ring & 6.6 & \cite{phoo92}\\
NGC\,4565 & companions/bridge & $\sim$1 & \cite{rup91}\\ 
NGC\,4826 & counter-rotating disk & a few & \cite{braun94}\\
NGC\,5457 (M\,101, Arp 26) & high-velocity complexes & 2 & \cite{hulst88, kam93}\\
NGC\,5635 & cloud & 2 & \cite{sag88}\\
NGC\,6946 & plume & $\gsim$1 & \cite{boo05b}\\
Milky Way & Magellanic Stream & 1.2 &	\cite{mat74, bruns05}\\
\\
NGC\,2865 & & & \cite{schi95} \\
NGC\,3656 (Arp 155) & & & \cite{bal96}\\
NGC\,4472 (Arp 134) & & & \cite{mcna94}\\
NGC\,5128 (Cen A)& & & \cite{schi94}\\

\noalign{\smallskip}\hline
\end{tabular}
\end{center}
$^a$ G.\ Gentile, private communication;
$^b$ T.\ Oosterloo, unpublished data.
\end{table}

\begin{figure*}
\centering
\includegraphics[width=\textwidth]{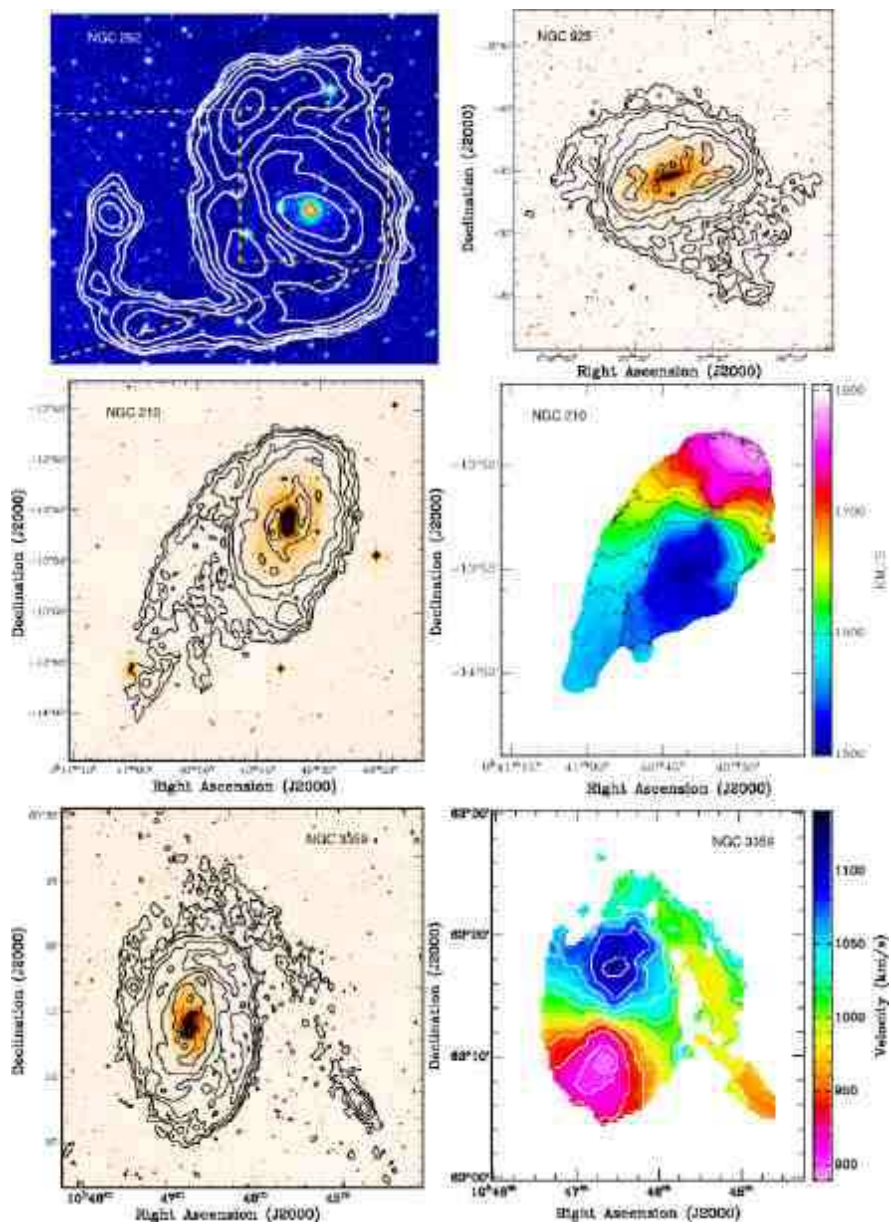}
\caption{Examples of galaxies showing signs of interactions/accretion.
In each panel the contours show the \hi\ density distribution 
superposed on the optical image. 
For NGC\,925 the levels are 5, 10, 20, 50, 100 $\times 10^{19} \cm$; 
for NGC\,210: 5, 10, 20, 50, 100, 200 $\times 10^{19} \cm$; 
for NGC\,3359: 10, 20, 50, 100, 200, 400 $\times 10^{19} \cm$. 
The \hi\ data for NGC\,262 are from Simkin et al (1987) \cite{sim87}.
The middle and bottom right panels show the velocity fields of NGC\,210
and NGC\,3359.
}
\label{f_interactions}
\end{figure*}

Prototypes of galaxies with interacting dwarf companions are 
NGC\,3359 (Fig.\ \ref{f_interactions}), NGC\,4565-4565A (Fig.\ \ref{f_n4565})
and NGC\,4027-4027A.
The companions have systemic velocities close to those of the main galaxy and
\hi\ masses less than 10 \% of the main galaxy. 
The \hi\ picture suggests the capture of a
gas-rich dwarf by a massive system, probably to be followed by tidal
disruption and accretion of the dwarf. 
This would bring in gas, stars and dark matter.

The Milky Way and the Magellanic Clouds are in this class of phenomena and 
the Magellanic Stream \cite{mat74,bruns05} is the gas component 
(about $1.2 \times 10^8 \mo$) probably destined to be accreted by our galaxy.
The \hi\ masses of LMC and SMC (4.4 and 4.0 $\times 10^8 \mo$, respectively) 
are of the same order as those of the companions discussed here.

The cases just mentioned probably represent early stages of the
interaction-accretion process. At later, more advanced stages, 
the victim may be no longer visible or not easy to  
be identified unambiguously. 
Examples are 
M\,101 (see \ref{m101}, Fig.\ \ref{f_m101}), 
NGC\,210 (see \ref{n210}, Fig.\ \ref{f_interactions}), NGC\,1023, NGC\,3310, 
NGC\,628 and Mkn\,348 (Fig.\ \ref{f_interactions}).  
NGC\,1023 is an S0 galaxy surrounded by a
clumpy and irregular \hi\ structure of 
$1.0\times10^9 \mo$ \cite{san84}, reminiscent of the
tails and bridges found in interacting multiple systems.
There are a few dwarf neighbours, one in particular on its eastern edge,
which might be merger relics.  
NGC\,3310 is a peculiar (Arp 217) Sbc starburst galaxy. 
Mulder et al. (1995) \cite{mul95}
and Kregel \& Sancisi (2001) \cite{kre01} have shown the presence of extended
\hi\ emission, which has a well developed two-tail structure. 
This must be an advanced merger 
that has either preserved the old disk of one of
the progenitors or, perhaps more likely, has led to the formation of a
new disk.  
Although the optical images of both NGC\,1023 and NGC\,3310,
as of several other objects in Table \ref{t_interactions} 
(e.g.\ M\,101, NGC\,925), already 
show some peculiarities, it is their \hi\ structure and
kinematics that fully reveal the ongoing mergers.
Other galaxies, such as NGC\,210, NGC\,628 and Mkn\,348, have a clean, 
regular optical image and only the \hi\ betrays a possible recent
accretion. For NGC\,628 this is indicated by the presence in its outer
parts of two giant high-velocity \hi\ complexes, which
are symmetrically placed with respect to the galaxy center. 
These complexes have \hi\ masses of about 10$^8\mo$, 
and maximum velocity excesses of 100 $\kms$ \cite{kam92}.
For Mkn 348 (NGC\,262) a probable past interaction
and gas accretion is suggested by the presence of an enormous
\hi\ envelope (176 kpc diameter \cite{heck82}) 
and a large tail-like extension (Fig.\ \ref{f_interactions}) \cite{sim87}.
In all cases, like these latter, where there is no optical victim visible, 
we may be dealing with the accretion of a dwarf galaxy or of intergalactic clouds.
In the case of a dwarf, the stellar component has either already fallen in and has 
been digested, or it has been totally disrupted and the stars are now 
scattered around. Deep optical imaging may be able to  
reveal them. Examples may be the unusual faint features around some galaxies 
reported by Malin and Hadley \cite{mal97} (see also below).

It is important to note that in all cases with no obvious interaction, a
careful study of the structure and kinematics of the
\hi\ is necessary to distinguish between
configurations that can be considered ``normal'' and configurations
that are definitely ``peculiar'' and point to a recent interaction and
infall. There are recognizable signatures in the \hi\ that 
make this distinction possible, but it is not always easy to
draw the line between effects due to the internal metabolism of the
galaxy and those due to the environment. As an example of this difficulty, 
it is interesting to consider lopsidedness, which 
affects spiral galaxies and seems to occur quite frequently 
(Section \ref{lopsidedness}).  
Should it be attributed to past interactions and
accretion events? This is not at all obvious and there may be other
explanations related to the intrinsic dynamics of the system ($m=$1 mode)
\cite{saha07}.

Recently, accretion of satellites  has also been 
revealed by studies of the distribution and kinematics of stars in
the halos of the Milky Way and of M\,31. 
The discovery of the Sgr Dwarf galaxy \cite{ibata94}
is regarded as proof that accretion is still
taking place. Since such minor merger remnants
retain information about their origin for a long time \cite{helmi00},
studies of the distribution and kinematics of ``stellar
streams'' can in principle be used to trace the merger history of the
Milky Way \cite{helmi01}.
Stellar streams have also been discovered in M\,31 \cite{ibata01,ferg02,mccon03}. 
Such events are more difficult to trace in more 
distant galaxies, where it is not possible to observe individual stars.
However, deep optical images of a number of spiral galaxies, 
such as NGC\,253, M\,83, M\,104, NGC\,2855, \cite{mal97}
and NGC\,5907 \cite{shang98}, do show unusual, faint features in their surroundings.
These galaxies do not have obvious interactions or companions (except NGC\,5907 
which has a nearby dwarf). To our knowledge, a clear association of these 
features with \hi\ has not been shown yet, except, perhaps, for the association
between the Orphan stream and complex A discussed 
by Belokurov et al.\ (2007) \cite{belo07}.
It would be interesting to have deep optical images for 
\hi\ systems like those illustrated in Fig.\ \ref{f_interactions}.
 
For the study of interactions and accretion, especially in the case of 
distant galaxies, the \hi\ has clear advantages. 
As has been shown for our galaxy and for nearby galaxies, 
\hi\ not only provides a direct measure of the accreting gas, 
but is also a powerful tracer of merger events. In particular, 
the \hi\  distributions and kinematics can be used for
modelling and for estimating timescales.
The improved sensitivity of modern synthesis radio telescopes 
brings within reach the detection of fainter and fainter \hi\ 
signatures of accretion events and we expect that new observations of nearby and
also more distant galaxies will reveal many more examples in the coming decade.

Neutral hydrogen found in early-type galaxies (E and S0) reveals a similar 
picture as illustrated above for spirals, indicating that the accretion 
phenomenon is probably playing an important role in all types of galaxies 
(see bottom Table \ref{t_interactions}).
Recent surveys of \hi\ in early-type galaxies  in the field 
\cite{oos07b,sadler01,sadler02, morganti06}
show that about 60$-$70\% of 
them have detectable amounts of \hi\ (detection limits $10^7 \mo$). 
Similar detection rates were found in E and S0 galaxies with optical 
fine structure, such as optical shells \cite{schi97, vgork97} (e.g.\ NGC\,5128 (Cen A) \cite{schi94} and NGC\,2865 \cite{schi95}), and also near ellipticals with dwarf companions, 
like NGC\,4472 \cite{mcna94} and NGC\,3656 \cite{bal96}.
About half of these galaxies do show apparently relaxed, gaseous disks in 
regular rotation. 
The other half have irregular \hi\ distributions suggesting accretion and 
minor mergers in progress \cite{schi97, oos07b}. 
It is also interesting to note that there are gas-rich ellipticals, such as NGC 4278, that contain a lot of \hi\ but have a purely old stellar population, indicating 
that gas accretion and star formation are not tightly correlated.

\subsection{Specific examples}

To further illustrate the processes of interaction with dwarf 
companions and of merger/accretion we describe a few examples in detail.

\subsubsection{M\,101}
\label{m101}

\begin{figure*}
\centering
\includegraphics[width=\textwidth]{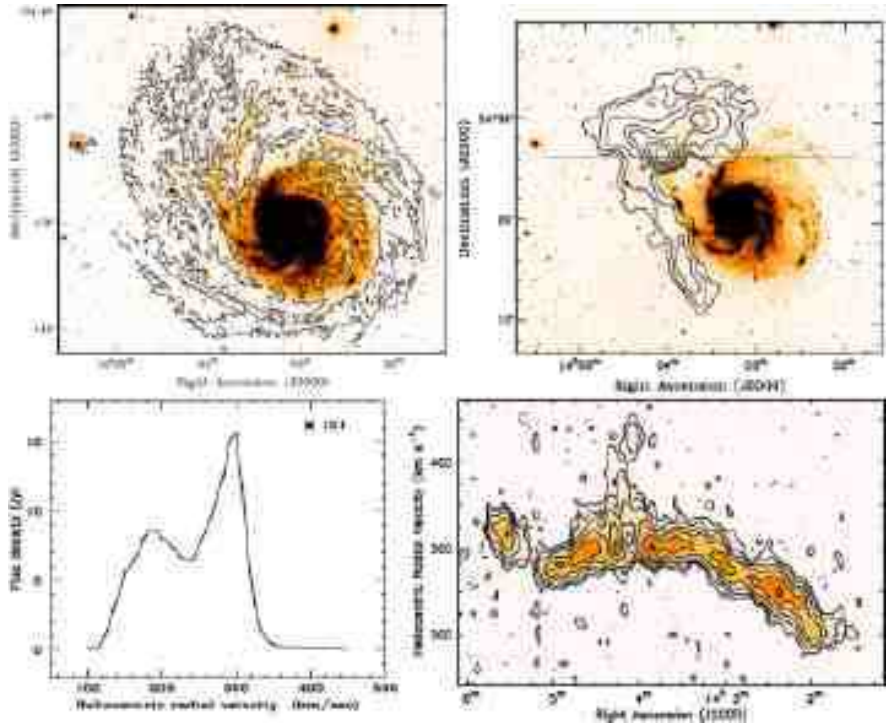}
\caption{ 
Top left: WSRT total \hi\ map for M\,101 (contours) overlaid on a DSS image.
Top right: high-velocity gas complex (contours) overlaid on the optical image.
Bottom left: global \hi\ profile.
Bottom right: Position-velocity diagram ($45''$ beam) at constant declination 
(see horizontal line in top right panel) showing the high-velocity \hi\ complex. 
The \hi\ data are from  \cite{kam93} (see also \cite{hulst88}).
}
\label{f_m101}
\end{figure*}

In M\,101 an \hi\ complex of about $2\times 10^8 \mo$ (Fig.\ \ref{f_m101}, 
top right panel)
has been found moving with velocities of up to $150 \kms$ with respect 
to the local disk and in correspondence with a large trough in the \hi\ layer  
(Fig.\ \ref{f_m101}, bottom right). 
It has been suggested that this is the result of a collision with a dwarf 
companion (not visible) or with a gas cloud complex which has gone through the 
\hi\ layer of M\,101 and has created the observed trough \cite{hulst88, kam93}. 
The high-velocity gas will eventually rain back down onto the M\,101 disk. 
It is interesting to note that M\,101 is a prototype lopsided galaxy
\cite{bal80} (Fig.\ \ref{f_m101}, top left; see Section \ref{lopsidedness}). 
The lopsidedness is also manifested by the global \hi\ profile (Fig.\ \ref{f_m101}, bottom left).

\subsubsection{NGC\,210}
\label{n210}

NGC\,210 is a good example of a galaxy that shows no indications 
of accretion or interaction in the optical, but where the \hi\ data give 
clear evidence for a recent merging event. 
Fig.\ \ref{f_interactions} (middle left panel) shows that 
NGC\,210 is a regular barred galaxy with well developed, symmetrical spiral 
arms. In the region of the optical disk, the \hi\ contours indicate a regular 
structure. In the outer parts, however, the \hi\ shows a long tail extending to the SE and containing about $6 \times10^8\mo$. This is 10\% of the total \hi\ mass of NGC\,210. 
The kinematics of the tail is similar to that of the gas disk, i.e.\ the tail 
is co-rotating with the disk, at similar projected velocities
(Fig.\ \ref{f_interactions}, middle right panel). 
Its orbital period is $1-2 \times 10^9$ years. 
This is about the time it will take for the asymmetric structure 
to wind up and disappear. 
Near the end of the tail, a small galaxy is visible in the optical. 
However, its redshift is known \cite{dacosta98}
and shows that it is  a background object at about ten times the distance of 
NGC\,210, excluding an interaction.
It is possible that the \hi\ tail is caused by a merger, 
although no signs of a victim are seen in optical data.

\subsubsection{NGC\,925}	
\label{n925}

Another example of a galaxy with an \hi\ tail suggesting an 
accretion event is NGC\,925. 
Fig.\ \ref{f_interactions} (top right) shows the \hi\ distribution 
in relation to the optical image  (see also \cite{pisano98}). 
The \hi\ properties are similar to those of NGC\,210, 
and it appears that also NGC\,925 has suffered an accretion event fairly recently. 
NGC\,925 has a tail of \hi\ extending to the South. 
The tail contains about $3\times 10^8\mo$ of \hi, or 5\% of the total \hi\ 
mass of NGC\,925. 
Also in NGC\,925 the kinematics of the gas in the tail 
is not too different from that of the gas in the galaxy, i.e.\ it appears to 
co-rotate at similar projected velocities. The winding up and disappearance of 
this structure may take about $0.5-1.0 \times 10^9$ yr.  
NGC\,925 is different from e.g.\ NGC\,210 in that in 
the optical \cite{pisano00} the galaxy 
is fraught with asymmetries, both morphological and kinematical,  
suggesting that the accreted object may have been relatively more 
massive than in NGC\,210.

\subsubsection{NGC\,3359}
\label{n3359}

NGC\,3359 is a nearby SBc  galaxy with a dynamical mass 
of $1.6 \times 10^{11} \mo$ \cite{broeils97}
and an \hi\ mass of $1.9 \times 10^{10} \mo$.
It has well developed spiral structure both in the optical
and in \hi. The observations \cite{kam94, hulst05}
have revealed the presence of an \hi\ companion and a
long tail/bridge connecting to the outer spiral structure of 
NGC\,3359 (bottom Fig.\ \ref{f_interactions}). 
The companion has an \hi\ mass of $4.4\times 10^{8} \mo$.
There is also evidence for an optical counterpart.
Together with the connecting structure the total 
\hi\ mass is about $9.6 \times 10^8 \mo$ or 5\% of 
the total \hi\ mass of NGC\,3359. 
The \hi\ image of the companion is distorted.
The velocity structure of the \hi\ companion and the 
connecting \hi\ (bottom right Fig.\ \ref{f_interactions})
fits in very well with the regular velocity field of NGC\,3359.  
The regularity of the velocities suggests that the process has been going
on slowly for at least about one rotational period (about 
$1.5 \times 10^9$ yr).

\subsubsection{NGC\,2985 (UGC\,5253)}	
\label{n2985}

\begin{figure*}
\centering
\includegraphics[width=\textwidth]{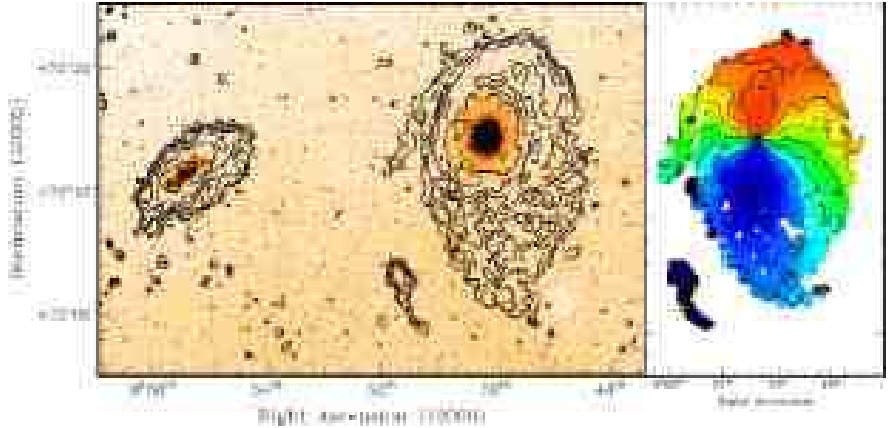}
\caption{Total \hi\ map of NGC 2985 (UGC\,5253) and its companions and 
velocity field.
On the left plot, the shade is the DSS and contours are \hi.
Contour levels are: 5, 10, 20, 50, 100, 200 $\times 10^{19} \cm$.}
\label{f_n2985}
\end{figure*}

Fig.\ \ref{f_n2985} shows the total \hi\ map (contours) for NGC\,2985 
(the galaxy on the right) and its surroundings overlaid on a DSS image. 
NGC\,2985 is a spiral galaxy at a distance of 18 Mpc. 
A number of features can be noted that are relevant here. 
First, the \hi\ distribution is very asymmetric. 
This may be due to an interaction with NGC 3027, the galaxy 20$'$ (120 kpc) 
east of NGC\,2985 and of similar redshift as NGC 2985 (velocity difference 
$\sim 250 \kms$). 
NGC\,3027  also has an asymmetric \hi\ distribution. 
Interestingly, the velocity field of NGC 2985 is, overall, fairly regular
and dominated by differential rotation (Fig.\ \ref{f_n2985}, right panel). 
It looks as if the passage of NGC\,3027 has caused a tidal displacement of the 
outer disk of NGC\,2985 with respect to the inner disk without destroying it 
and making NGC 2985 very lopsided in appearance.
The timescales for the disappearance of the asymmetry are about 
$1.5-3 \times 10^9$ yr.

Another interesting feature is the small galaxy SE of NGC 2985. 
In the optical, there is a faint, low surface brightness object, coincident 
with the peak of the \hi. The \hi\ map shows a comet-like structure suggesting that this small galaxy is interacting with NGC 2985, 
losing part of the \hi\ in the process. 
It is likely that this is an accretion of a small galaxy that we observe 
at an early stage of the process.  
NGC 2985 is very \hi\ rich, its \hi\ mass is $1.1 \times 10^{10} \mo$. 
The \hi\ mass of the small companion is about 3\% of that of NGC 2985, or 
$3.6 \times 10^8 \mo$, its systemic velocity is $185 \kms$ blue-shifted
with respect to that of NGC\,2985.

\subsubsection{NGC\,4565}
\label{n4565}

\begin{figure*}
\centering
\includegraphics[width=\textwidth]{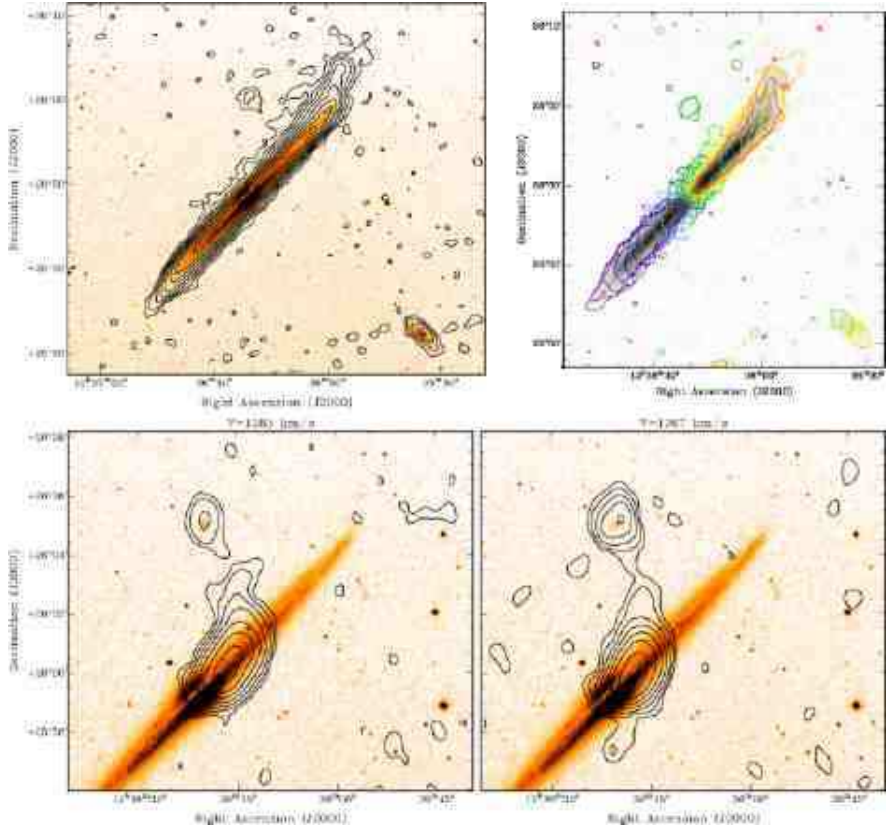}
\caption{Top left: \hi\ map of NGC 4565 at a resolution of 30$''$ 
superposed on the DSS image. 
Contours are 2, 4, 8, 16, 32 and 64 $\times 10^{20}$ cm$^{-2}$.
Top right: outer contours of the \hi\ emission in individual channels 
(from blue to red) superposed on the total \hi\ density map.
Bottom panels: \hi\ channel maps at two representative velocities 
superposed on the DSS image of NGC 4565. They
clearly show the interaction between NGC 4565 and its small companion. 
Contours are 1, 2, 4, 8, 16, 32 and 64 mJy/beam.
}
\label{f_n4565}
\end{figure*}

NGC\,4565 is a large edge-on galaxy of Hubble type Sb 
with a dwarf companion 6$'$ ($\sim$30 kpc) to the north of the center, 
F378-0021557, which has $7.4 \times 10^{7} \mo$
of \hi\ compared to an \hi\  mass of $2.0 \times 10^{10} \mo$ for NGC\,4565 (using
a distance of 17 Mpc) \cite{hulst05}. 
An \hi\ detection of this same companion 
has also been reported (with the name NGC\,4565A) by Rupen (1991) \cite{rup91}. 
Another companion, NGC\,4562, somewhat more massive in \hi\
($2.5 \times 10^{8} \mo$) and brighter optically is located 
15$'$ ($\sim$75 kpc) to the south-west of the center of NGC\,4565. The
\hi\ distribution, derived (by us) from a new sensitive observation 
with the Westerbork Synthesis Radio Telescope (WSRT) 
by Dahlem et al.\ (2005) \cite{dahlem05}, is shown in Figure \ref{f_n4565}
(top left panel) superposed on the DSS. 
The asymmetric warp is clearly visible.

Individual channel maps show  low surface brightness \hi\ 
emission to the north of the
centre, in the direction of the faint companion F378-0021557. 
The \hi\ emission from NGC\,4565 in the velocity range from 1250 to
1290 km s$^{-1}$ (close to the velocity of F378-0021557 and to the
systemic velocity, 1230 km s$^{-1}$, of NGC\,4565) clearly shows
distortions above the plane pointing towards the companion. 
This is seen in the map (Fig.\ \ref{f_n4565}, top right) showing the \hi\ 
velocity structure and in the bottom panels of Figure \ref{f_n4565} 
which show two channel maps chosen at velocities in this range. 
In these maps one can clearly see the \hi\ layer bending towards
F378-0021557, suggesting a connection with the dwarf and a strong
disturbance in the \hi\ disk of NGC\,4565. 
While there seems to be little doubt that this bending of the 
\hi\ layer is due to the interaction with the companion, 
it is not clear whether there is any relationship with the warp.
It is likely that, eventually, the companion will merge with NGC\,4565.

\subsubsection{NGC\,6946}
\label{n6946}

\begin{figure*}
\centering
\includegraphics[width=\textwidth]{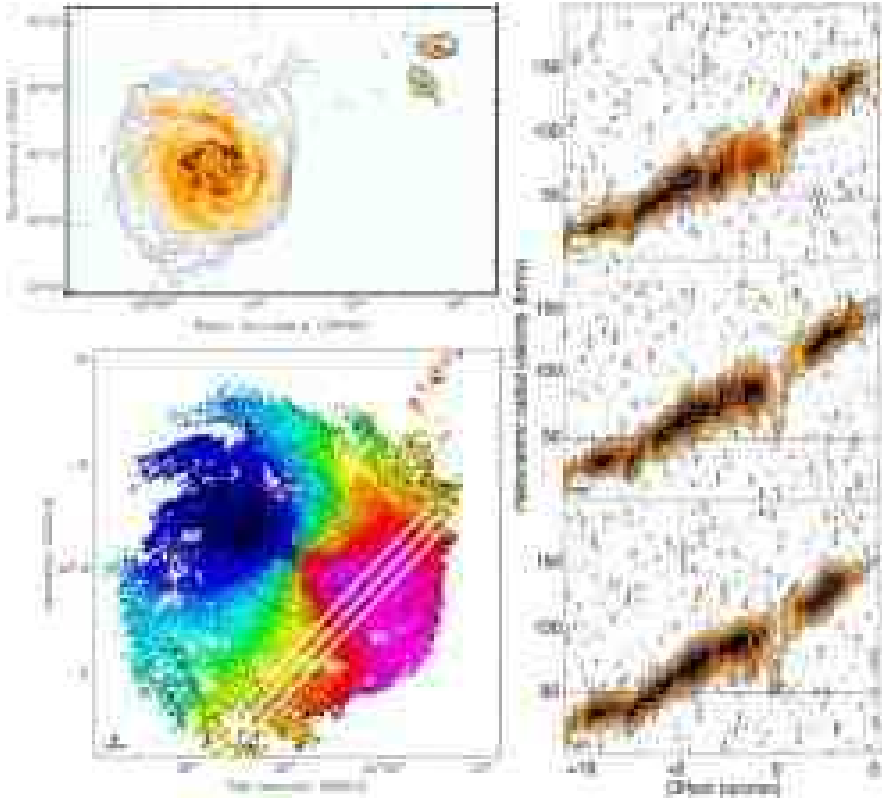}
\caption{Top left:  total \hi\ map of NGC 6946 and two companions. 
The map has been corrected for primary beam attenuation. 
The shading shows the high resolution (13$''$) \hi\ map,  
the contours (1.25, 2.5, 5, 10, and 20 $\times 10 ^{19} \cm$) 
show the low resolution (60$''$) \hi\ distribution in the outer parts.
Bottom left: velocity field at 22$''$ resolution 
The iso-velocity contours are separated by 10$\kms$ running 
from −70 (blue) to 150$\kms$. 
The small ellipse indicates the size and orientation of the optical disk 
(R$_{25}$). 
Right panels: position-velocity plots along the three white lines overlaid
in the velocity field in the bottom left panel 
\cite{boo07}.} 
\label{f_n6946large}
\end{figure*}

Fig.\ \ref{f_n6946large} (top left) shows a total \hi\ image of
NGC\,6946  down to column density levels of $1.3 
\times 10^{19}$ cm$^{-2}$ \cite{boo07}.
At about 36$'$ ($\sim$70 kpc) to the north-west 
there are two small companions. 
An intruiging feature in the \hi\ image of NGC\,6946 is the plume 
to the north-west. This is 20 kpc long and blends smoothly (also kinematically) 
with the \hi\ disk of NGC\,6946. 
A lower limit to its \hi\ mass is $7.5 \times 10^7 \mo$. 
This is similar to the \hi\ masses of the companions ($1.2 \times 10^8 \mo$
 and $8.8 \times 10^7 \mo$). 
Although the plume is in the same range of radial velocities as the two 
companion galaxies there is no detected connection with them. 
As in many other cases of peculiar features found around galaxies,  
it is not clear what its origin is and whether this is new material 
arriving from outside or whether it is the result of a tidal perturbation 
of the outer gaseous layer of NGC\,6946.
It might even be the accretion, seen at an advanced stage, of a third 
companion galaxy similar in mass to the other two.

In NGC\,6946 there is further evidence pointing to a possible infall of material 
from outside. In the first place there are the striking spiral arms in the outer 
\hi\ disk (see section \ref{extra6946}), well outside the bright stellar disk 
and the regions of star formation. 
Secondly, in the arm-interarm regions and following the spiral arms, there are 
strong velocity wiggles. These wiggles, clearly visible in the velocity field (Fig.\ \ref{f_n6946large}, bottom left),
are abrupt velocity deviations from circular motion reaching amplitudes 
of about 50$\kms$ (see p-v plots in the right panels of Fig.\ \ref{f_n6946large}). 
The corresponding troughs in the \hi\ density distribution 
seem to indicate that the \hi\ layer has been 
punched by infalling clumps of material. 
The picture is similar to that seen in M\,101 
and attributed to collision with intervening gas clouds.

The high resolution \hi\ image of NGC\,6946 is shown in Fig. \ref{f_n6946},
where it is compared with optical images, and the extra-planar \hi\ 
is discussed below in subsection \ref{extra6946}.

\subsection{Frequency of accretion events}
\label{frequency}

How frequent are the interactions of galaxies with small companions 
and what is the fraction of galaxies with peculiar morphology and 
kinematics? What is the rate of accretion expected from all these presumed 
minor mergers?  
In the past years, a large number of galaxies have been mapped in
\hi\ with the WSRT, the Very Large Array (VLA), the Australia Telescope  
Compact Array (ATCA) and the Arecibo radio telescope. 
A first estimate made on the basis of about one hundred
galaxies led to the conclusion \cite{san92} that at least 25\%
of field galaxies show signs of either present or recent tidal 
interactions. The incompleteness and inhomogeneity of the
sample examined made such an estimate rather uncertain. 
An \hi\ survey carried out for a magnitude and volume limited sample of 
galaxies from the Ursa Major cluster  \cite{ver01}  provides more solid 
statistical evidence on the frequency of tidal interactions and of accretion 
phenomena. 
This cluster differs from Virgo or Coma type clusters. 
It has a low velocity dispersion and long crossing time, comparable to the Hubble time. 
It has no central concentration and no detected X-ray emission and
the sample is dominated by late-type systems. 
It can be considered, therefore, representative for a galaxy population in the field.  
Out of the 40 galaxies mapped in \hi, about 10 show
clear signs of interactions with small companions or have peculiar
structures. About half of the sample galaxies show asymmetries in their 
kinematics or in the \hi\ density distribution. 
A larger sample of galaxies is the one provided by WHISP \cite{hulst01}.
About 25\% of 300 spirals and irregulars show evidence of minor interactions.

In conclusion, the available evidence from \hi\ 
observations indicates that at least 25\% of field galaxies are
undergoing now or have undergone in the recent past some kind of tidal
interaction. The lifetimes of the observed features are typically $\sim$1 Gyr. 
If lumps of gas with \hi\ masses of order 10$^{8-9} \mo$ (as
indicated by the 21 cm observations) are accreted at a rate of 1 per
10$^9$ yr, the mean accretion rate for the gas would be around $0.1-0.2 \moyr$.
This is certainly a lower limit for 
gas accretion as a fraction of the \hi\ involved in the interaction  
may be undetected and neither ionized hydrogen nor helium 
have been considered.
Furthermore, the number of past interactions and mergers may be higher.
Indeed, if one is willing to accept also the lopsided structure and kinematics 
as evidence (see below), as also proposed in optical studies \cite{zar97}, 
then the conclusion would be that more than 50\% of present day 
galaxies have been through one or more merger events in a recent past.
In such a case the accretion rate would be difficult to estimate, but 
it could be considerably higher than the values given above.

\section{Extra-planar \hi}
\label{extraplanar}

\subsection{Galaxies with \hi\ halos}
\label{halos}

\begin{figure*}
\centering
\includegraphics[width=0.75\textwidth]{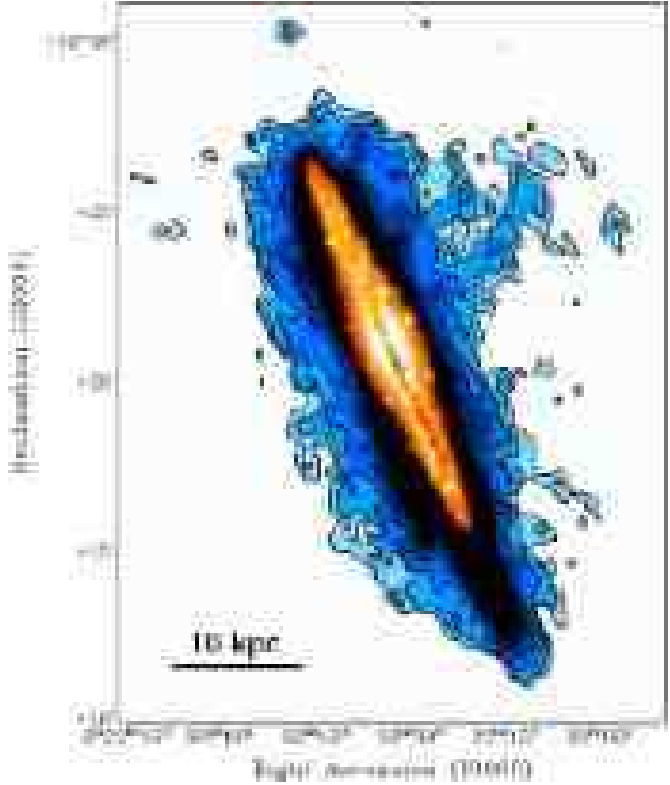}
\caption{Optical DSS image (red) and total \hi\ map (contours + blue shade) 
of the edge-on galaxy NGC 891. \hi\ contours are at 1, 2, 4, 8, 16 $\times 10^{19}\cm$ \cite{oos07}.
The beam size is 25$''=$1.1 kpc. 
}
\label{f_n891}
\end{figure*}

The presence of cold gas in the halo region of disk galaxies (extra-planar gas) 
is well established. 
For a small number of systems (Table \ref{t_extraplanar}) seen at various inclination angles and, in particular, for the edge-on galaxy  NGC\,891 
(see Fig.\ \ref{f_n891}) there is now detailed information on the \hi\ structure and kinematics. This has been obtained from very deep  observations 
with the WSRT and the VLA. 
Also the HVCs of our galaxy (at least the largest) are now regarded, 
as a result of the recent distance determinations \cite{wak07,wak08}, 
 as a Galactic halo population.
The analogy between the Galactic HVCs and the high-velocity \hi\ in external 
galaxies has been discussed by Oosterloo (2004) \cite{oos04}.

What is the origin of the extra-planar gas?
Undoubtedly, a large fraction has originated from the disk as an effect of star formation. 
The mechanism is that of a ``galactic fountain'' in which hot gas rises into the halo, condenses into cold clouds and returns to the disk (first suggested by Shapiro \& Field 1976 \cite{sha76}; see also \cite{breg80}).
There are various indications from the \hi\ observations that extra-planar gas 
is indeed driven by star formation. One is the distribution of the extra-planar \hi\ in NGC\,891, which is concentrated very close to the star-forming disk 
(see \ref{n891}). Another is the remarkable concentration of the majority of the high-velocity clouds in NGC\,6946 (see \ref{extra6946}) in the direction of the bright inner disk.

However, there is also evidence that a fraction of the extra-planar gas 
must be infall from intergalactic space.
This is indicated primarily by the low metallicity of the HVCs, which points 
directly to an external origin: infall of pristine gas clouds or gas-rich dwarf companions. 
Similarly, in other galaxies, such infall is supported by the presence of huge \hi\ filaments and clouds with peculiar motions. 
An argument in favour of accretion may also come from 
the large-scale kinematics of the halo gas, which is characterized by 
rotational velocity gradients along the $z$ direction 
and a global inflow motion.
It has been suggested that such kinematics can be explained by the interaction
between the fountain gas and infalling gas, which carries low angular momentum \cite{fb08}.
However, the gradients by themselves may not necessarily require infall (e.g.\ \cite{barn06}).

Here, we give a short review of the main, relevant observational results  and 
we end with estimates of accretion rates.
We describe the observations for the best cases known. For the edge-on galaxies 
(e.g. NGC\,891) the extent and structure of the halo are observed directly and 
the rotational velocities are measured at various distances from the plane. 
In galaxies at lower inclination angles, such as NGC\,2403 and NGC\,6946, 
the presence of extra-planar gas is inferred from the observed anomalous kinematics (high velocities, slow rotation). 
Indeed, it is this anomalous kinematics that is used to 
separate the extra-planar gas from the thin disk. 
In particular, in NGC\,6946, which is closer to face-on, one sees a large 
number of clouds with high velocities, i.e.\ with large velocity deviations 
from circular motion, which can be unambiguosly separated from the 
differentially rotating disk.
It should be emphasized  that all these galaxies are not undergoing strong 
gravitational interactions and, therefore, the observed high-velocity 
structures are not tidal features.

\begin{table}[t]
\caption{Extra-planar gas in spiral galaxies}
\begin{center}
\label{t_extraplanar}       % Give a unique label
\begin{tabular}{lcccccc}
\hline\noalign{\smallskip}
Galaxy & Type & incl & v$_{\rm flat}$ & M$_{\rm HI_{\rm halo}}$ & $\frac{M_{\rm HI_{\rm halo}}}{M_{\rm HI_{\rm tot}}}$ & Ref.\ \\
       &      & (deg)& ($\kms$)       & ($10^8 \mo$)        & (\%)                     & \\
\tableheadseprule\noalign{\smallskip}
Milky Way & Sb & -  & 220 & $>0.2$ & $>1^a$ & \cite{wak07}\\
M\,31     & Sb  & 77 & 226 & $>0.3$ & $>1$ & \cite{thi04}\\
NGC\,891  & Sb  & 90 & 230 & 12   & 30 & \cite{oos07}\\
NGC\,6946 & Scd & 38 & 175 & $>$2.9 & $>$4 & \cite{boo05b}\\
NGC\,4559 & Scd & 67 & 120 & 5.9 & 11 & \cite{barb05}\\
NGC\,2403 & Scd & 63 & 130 & 3   & 10 & \cite{fra02a}\\
UGC\,7321 & Sd  & 88 & 110 & $\gsim 0.1$ & $\gsim$1 & \cite{mat03}\\
NGC\,2613 & Sb  & $\sim$80& $\sim$300 & 4.4$^b$ & 5 &  \cite{cha01}\\
NGC\,253  & Sc  & $\sim$75& $\sim$185 & 0.8 & 3 &  \cite{boo05a}\\
\noalign{\smallskip}\hline
\end{tabular}
\end{center}
{\footnotesize $^a$ Only HVCs (IVCs not included); 
$^b$ from sum of the various extra-planar clouds.}
\end{table}

\subsubsection{Milky Way}
\label{MW}

The High-Velocity Clouds (HVCs) of neutral hydrogen in our galaxy \cite{wak97}
(see also recent review by Van Woerden et al.\ (2004) \cite{woerden04})
(Fig.7, top panel) have been considered since their discovery 
as possible direct evidence for infalling gas. The lack of information on their distances and therefore on their masses has been, however, a major obstacle. It has even been proposed that they are a 
population of clouds in the Local Group of Galaxies \cite{bli99}.
Recently it has been possible, through the study of absorption lines 
in the spectra of stars, to set distance brackets on some of the 
largest and more massive clouds like complex A (8$-$10 kpc, \cite{woerwak04})
and complex C (3.7$-$11.2 kpc, \cite{wak07}). For this latter, the conclusion 
is that it must be located high above the Galactic plane (z=3$-$9 kpc) 
and at a Galactocentric radius R$<$14 kpc. Its mass estimate is 
$3-14 \times 10^6 \mo$, its metallicity is 0.15 times solar.
Such low metallicities, also found for other HVCs \cite{woerwak04},
establish that most of this gas must be fairly pristine and hence 
of external origin (from  gas-rich dwarfs or 
intergalactic clouds) and not ejected from the disk.  
The mass inflow is estimated at 0.1$-0.25 \moyr$, including 
ionized hydrogen and a 40\% contribution from helium. 
The  masses and sizes of these HVC complexes 
are similar to those of the filaments found in NGC\,891 
and in NGC\,2403 and reported below.
The Intermediate-Velocity Clouds (IVCs) form a component 
closer to the Galactic layer. 
It is now clear that IVCs and HVCs are a halo population of the 
Milky Way, analogous to the gaseous halos found in external galaxies. 
Indeed, seen from outside, the halo of our galaxy (IVCs included) 
might well look like the halos of NGC\,891 and NGC\,2403.

\begin{figure*}
\centering
\includegraphics[width=\textwidth]{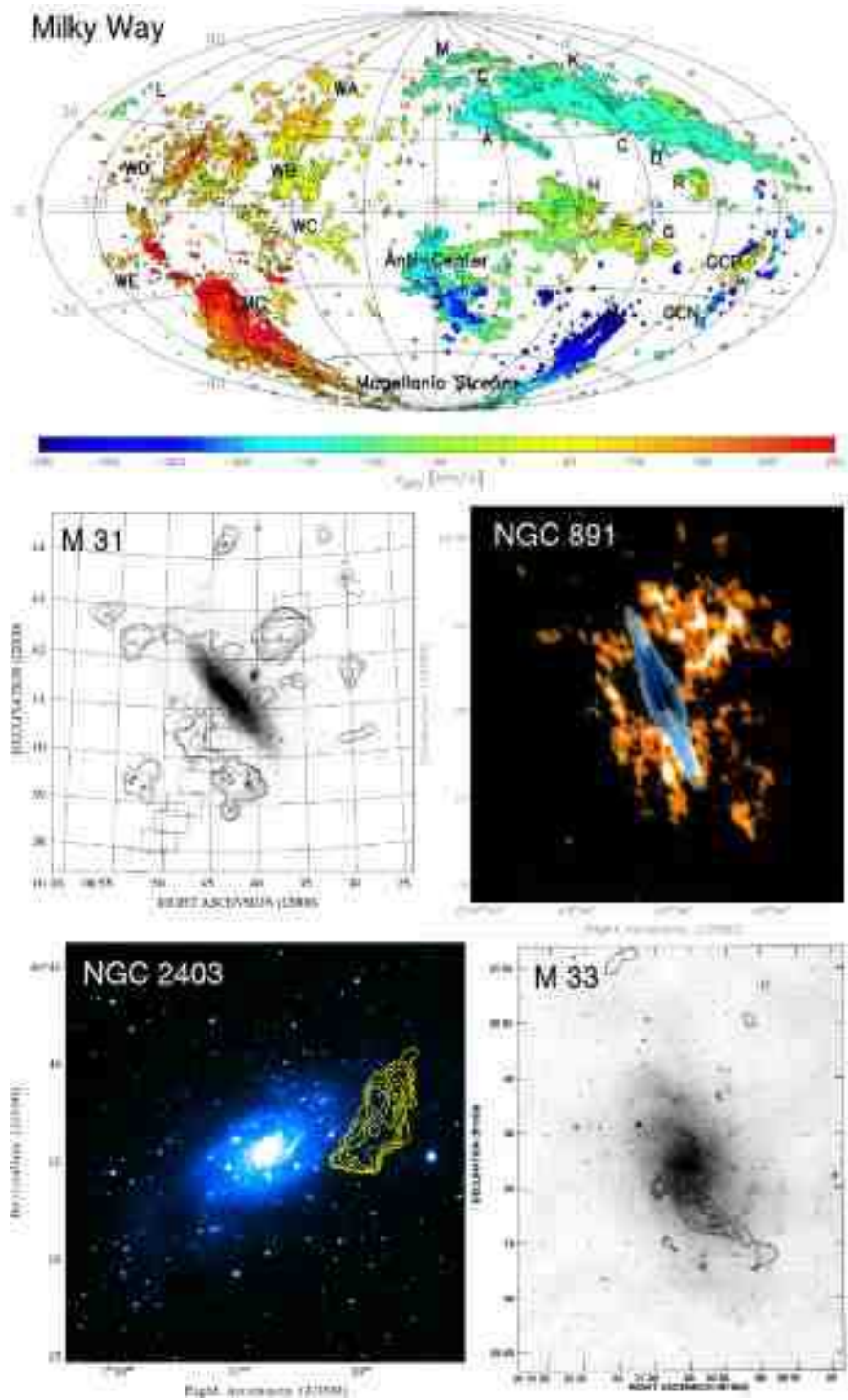}
\caption{Examples of extra-planar gas complexes around nearby galaxies.
From top: High-Velocity Clouds in the Milky Way, extra-planar gas 
features in M\,31 and NGC\,891. 
The bottom panels show two massive \hi\ filaments in NGC\,2403 and M\,33.}
\label{f_clouds}
\end{figure*}

\subsubsection{M\,31}
\label{m31}

\hi\ observations with the Green Bank Telescope (GBT) by Thilker et al. (2004)
\cite{thi04}
have revealed a population of faint \hi\ clouds (at least 20) surrounding M\,31
within 50 kpc of its disk (Fig.\ \ref{f_clouds}, middle left)
and with radial velocities comparable to those of 
the outer disk rotation. The masses of these clouds are in the range 
10$^5-10^7 \mo$. In addition, a filamentary component of at least 	
30 kpc extent is concentrated at the M\,31 systemic velocity. 
Thilker et al.\ argue that all this gas is associated with M\,31. 
The total amount of \hi\ for the halo cloud population within the GBT field 
 is estimated to be $\sim3-4 \times 10^7 \mo$, 
which is only 1\% of the mass of the \hi\ disk of M\,31. 
This is probably a lower limit.
For the origin of this M\,31 halo Thilker et al.\ suggest various possibilities:
a Local Group cooling flow, tidal debris from recent mergers or 
the gaseous counterpart of low-mass dark matter halos.

\subsubsection{NGC\,891}
\label{n891}
	
The nearby edge-on galaxy NGC\,891 has been observed in \hi\ a number 
of times over the past three decades with increasing sensitivity 
\cite{san79, rup91, swa97, oos07}.
With the sensitivity improvement by about a factor of 50 
from the first to the latest observations (Fig.\ \ref{f_n891}, 
see also \cite{oos07}) it has been possible to trace 
the \hi\ emission in the vertical direction to 22 kpc from the plane,
whereas the size of the \hi\ disk (as measured along the major axis) 
has remained unchanged \cite{oos07}.
Radially, the halo extends to the end of the disk on the N-E side but stops 
earlier on the S-W side where the disk is more extended. This is an 
indication that the main bulk of the halo is closely connected to the 
inner disk of NGC\,891 where star formation is higher.
The mass of the \hi\ halo is about $1.2 \times 10^9 \mo$, $\sim$30\% of 
the total \hi\ mass.
Its kinematics is characterized by a slower rotation with respect to the 
disk, with a vertical gradient of $-15 \kmskpc$ \cite{fra05}.
The same gradient has been found also in the ionized gas \cite{hea06}.
In addition there is, at velocities close to systemic, a remarkable 
filament extending up to 22 kpc from 
the plane and containing $1.6 \times 10^7 \mo$. 
There are also other structures and in particular some clouds with 
anomalous (counter-rotating) velocities and masses of $1-3 \times 10^6 \mo$.

In Fig.\ \ref{f_clouds} we show the optical picture of NGC 891 surrounded by the \hi\ left  after subtraction of a ``normal'' (symmetrical, smooth, regular) disk and halo component (models from Oosterloo et al. 2007, Fig.\ 14 \cite{oos07}). 
The features seen here, some of them (e.g.\ the filament) recognizable 
in Fig.\ \ref{f_n891}, represent the \hi\ around NGC\,891 which is the most peculiar for its location and kinematics. Their total mass is about 
$1 \times 10^8 \mo$. We believe that these features may be regarded as the analogue of the HVCs in the Milky Way.

NGC\,891 has also a small, gas-rich companion (UGC 1807) at about 80 kpc 
(projected distance) and 100 $\kms$ higher radial velocity and with about one tenth of the total mass of NGC\,891.

\subsubsection{NGC\,2403}

\begin{figure*}
\centering
\includegraphics[width=\textwidth]{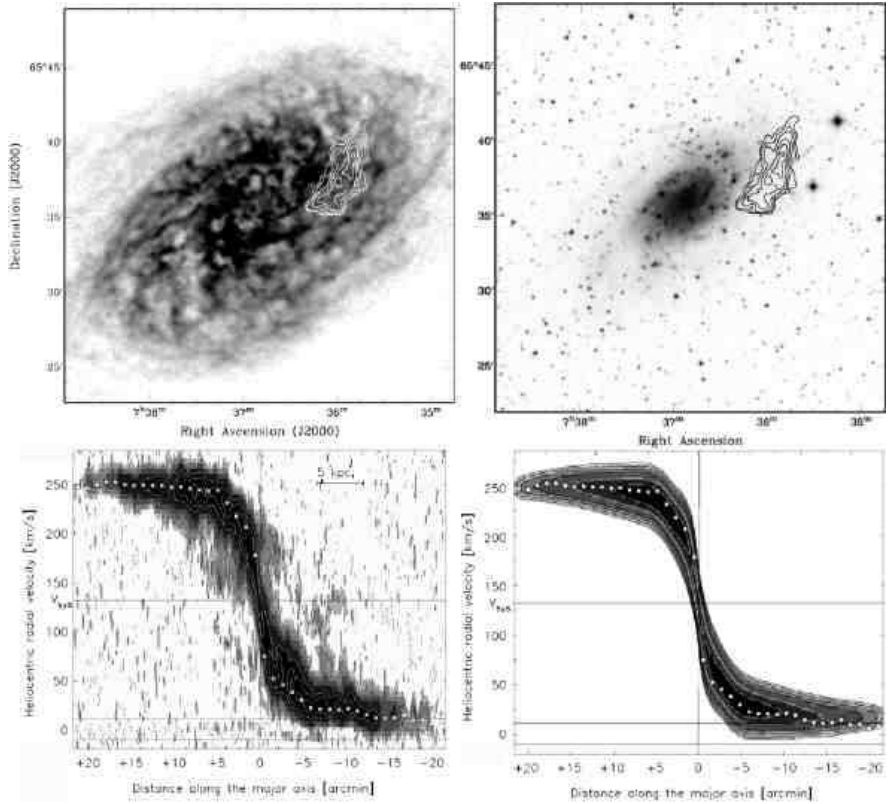}
\caption{Top panels: total \hi\ (VLA) map (left) and optical DSS image 
(right), on the same scale, for the spiral galaxy NGC\,2403.
The contours show a 8-kpc long \hi\ filament kinematically decoupled
from the bulk of the disk gas.
Bottom panels: comparison between the \hi\ position-velocity plot along the major axis of NGC\,2403 and the prediction 
for a thin disk model (right panel).
The gas at anomalous velocities, not visible in the model, is extra-planar
gas \cite{fra02a}.
}
\label{f_n2403}
\end{figure*}

For other galaxies seen at intermediate inclinations 
(NGC\,2403 \cite{scha00, fra02a} and NGC\,4559 \cite{barb05}) 
or more ``face-on''  (M 101 and NGC\,6946) the presence of gaseous 
halos has been inferred from the anomalous velocities.
In NGC\,2403 and NGC\,4559 the signature is 
the asymmetric velocity shape of the \hi\ 
line profiles (Fig.\ \ref{f_n2403}, bottom left). 
The deep \hi\ observations of NGC\,2403 with the VLA and a
careful 3-D modelling \cite{fra01} have led to conclusions 
very similar to those  obtained for the edge-on NGC\,891. 
Also in NGC\,2403 there is a vertically extended (a few kpc) 
gas component with lagging rotation. 
It has a total mass of about $3 \times 10^8 \mo$ ($\sim$10\% of the total 
\hi\ mass) and contains long filamentary structures. 
In addition to its lagging rotation, the halo gas has a large-scale inflow 
motion of about 15 $\kms$ toward the center of the galaxy.

The most striking filament (Fig.\ \ref{f_n2403} upper right, and 
Fig.\ \ref{f_clouds} bottom left) has coherent (narrow) velocity 
structure (with radial velocities close to systemic), 
it is 8 kpc long and  contains about $1 \times 10^7 \mo$. It is 
seen projected on the outer parts of NGC\,2403, beyond the bright optical disk. 
This is similar to the filament found in the halo of NGC\,891
and to the one found in M\,33 (Fig.\ \ref{f_clouds}, bottom right) 
(J.M. van der Hulst, unpublished data).
It is also close in size and mass to the largest Milky Way's HVCs 
like complex C.
Such filaments are the most remarkable and extreme structures 
found in the halo regions of these galaxies (Fig.\ \ref{f_clouds}). 
They may be quite common among spirals and be direct 
evidence of cold gas accretion from intergalactic space (see \ref{extragas_accretion}).

\subsubsection{NGC\,6946}
\label{extra6946}

NGC\,6946 is a bright, nearby spiral galaxy of Hubble type Scd
(inclination angle 38$^{\circ}$) which has been studied in \hi\ 
several times \cite{rog73, tac86, boul92, kam93}.
It was in this galaxy that Kamphuis and Sancisi (1993) \cite{kam93}
found evidence for an anomalous-velocity \hi\ component which
they interpreted as an outflow of gas from the disk into the halo as a
result of stellar winds and supernova explosions. A more recent, 
very sensitive and detailed study of the anomalous \hi\ and of 
the \hi\ disk has been carried out by Boomsma 
(2007) \cite{boo07} with the WSRT. 
The density distribution in the \hi\ disk is 
characterized by the presence of a large number of holes of diameters 
up to 2 kpc. The average amount of \hi\ missing from each hole is about 
$1 \times 10^7 \mo$, the total amount is $1.1 \times 10^9 \mo$.
Widespread high-velocity gas has been detected amounting to  a total 
of $2.9 \times 10^8 \mo$ (4\% of the total \hi\ mass) and deviating 
by $>50 \kms$ from local disk rotation. 
This gas is made of clumps and filaments of various sizes 
and most of it is seen projected against the H$\alpha$
bright inner disk of NGC 6946, which suggests a close relationship 
with the regions of star formation. 
A large part of it must be in the halo of NGC 6946, although its 
distance from the plane is not known. It follows the galaxy's  
differential rotation but there are clear indications that it is 
rotating more slowly than the gas in the disk. Overall, the picture 
is very similar to that of NGC 2403, and the conclusion reached by 
Boomsma (2007) \cite{boo07} is that the disk of NGC 6946 is surrounded 
by a lagging \hi\ halo similar to the halos found for NGC 891 and NGC 2403.
The outer parts of the disk of NGC\,6946 are discussed in subsection \ref{n6946}.

\subsubsection{Other galaxies}

There is, in addition to the galaxies described above, a number of other 
objects in which extra-planar gas (or traces of it) has been found. 
Some of these are Low Surface Brightness (LSB) galaxies, others 
have high surface brightness as those just described above or are even starbursts. 

Matthews and Wood (2003) \cite{mat03} 
find evidence for vertically extended \hi\ emission 
up to 2.4 kpc in the edge-on, superthin LSB galaxy UGC 7321.
They find tentative evidence that the vertically extended gas has declining 
rotational velocity as a function of $z$. 
They estimate the \hi\ mass of the 
halo (above $\sim$1.4 kpc) to be $\sim 1 \times 10^7 \mo$.
This value is approximately 15 times lower than the fractional \hi\ content 
(ratio between gas above 1.4 kpc and gas in the disk)  at comparable $z$-heights 
in NGC\,891.
In a similar, superthin LSB galaxy, IC\,2233, Matthews and Uson (2007) 
\cite{mat08} report a component of ``anomalous'' extra-planar gas.
Also in the LSB galaxy NGC\,4395 (see Section \ref{lopsidedness}), Heald \& Oosterloo 
(2008) \cite{heald08} find \hi\ cloud complexes with anomalous 
velocities, and presumably located in the halo, with masses of a few 
$\times 10^6 \mo$ each and, in total, about 5\% of the 
 \hi\ content of that galaxy. 

In the starburst galaxy NGC 253, Boomsma et al.\ (2005) \cite{boo05a} 
find extra-planar \hi\ concentrations reaching as high as 12 kpc above the disk. 
The gas seems to be lagging in rotation and has a total \hi\ mass of
$8 \times 10 ^7 \mo$. 
In the spiral galaxy NGC\,4559, Barbieri et al.\ (2005) \cite{barb05} 
find  extra-planar \hi\ with properties similar to those found for NGC\,2403.
Several high-latitude \hi\ features are observed in NGC\,5775 \cite{lee01}
and NGC 2613 \cite{cha01, irw03}; however, in the first case there is evidence
for a strong interaction with the companion galaxy NGC\,5774 \cite{irw94}.

\subsection{Accretion rate}
\label{extragas_accretion}

The fraction of \hi\ present in the halo of spiral galaxies seems to vary 
considerably from galaxy to galaxy (see Table \ref{t_extraplanar}). 
However, only for a few cases there are good estimates of the amounts of 
\hi\ gas in the halo.
These range from  about $3 \times 10^8 \mo$ for NGC\,2403 ($\sim$10\% of the total \hi) \cite{fra02a} to $1.2 \times 10^9 \mo$ (about 30\% of total) found for NGC\,891 \cite{oos07}. In our galaxy, the HVCs probably are only a small 
fraction of the extra-planar \hi\ and should be regarded as an ``extreme'' population. Clearly, for a better estimate and a proper comparison 
with external galaxies such as NGC\,891, one 
should include  the IVCs together with the HVCs.

To date, the number of objects studied with sufficient sensitivity and angular 
resolution is very limited and it is 
not possible yet to investigate possible dependencies on morphology, 
luminosity, surface brightness or star formation activity.
However, concerning the origin of extra-planar gas, there seems to be little 
doubt that, in high surface brightness galaxies, galactic fountains are 
responsible for a large part of it; but accretion from IGM is also taking place. 
Low surface brightness galaxies are more intriguing:
galactic fountains, if present at all, are expected to play a less important 
role and yet the observations of the LSB galaxies UGC\,7321, IC\,2233 and 
NGC\,4395 do reveal the presence of some extra-planar gas. 
Unfortunately most of these 
observations are not deep enough to trace the halo emission to large heights 
and to obtain a good estimate of the extra-planar \hi\ mass.

Accretion rates for extra-planar gas have been estimated using different
techniques.
Values for the gas accretion rate around 0.1 to $0.25 \moyr$ have been reported 
for the HVCs of our galaxy \cite{wak07}. 
These include ionized gas and helium.
The observations of some of the galaxies discussed here (NGC\,891, NGC\,2403 and M\,33) 
have revealed the presence of filamentary structures in their halos which are, 
like the HVCs, most likely of external origin. 
The main argument in favour of this interpretation 
is their high kinetic energy requirement (of order $10^4-10^5$ supernovae).
Moreover, some gas complexes are observed at forbidden (``counter-rotating'') 
velocities (see e.g.\ \cite{oos07}), which suggests material not dynamically
linked to the galaxy disk. 
One expects that the metallicities (not known) of all 
these filaments and gas complexes are as low as those found for the HVCs.
If we assume that all these features are accreted gas, we get values
for the ``visible'' accretion rate of $0.1-0.2 \moyr$ (\hi\ only), 
similar to the values obtained for the HVCs of the Milky Way.
This is of the order of 10\% the SFR in spirals like NGC\,891 and NGC\,2403, 
and also in the Milky Way.

The actual accretion rate may, however, be significantly higher and reach 
values approximately equal to the SFR, as a large fraction of the accreting gas 
may already be mixed with the fountain gas coming from the disk.
Such mixing could solve the problem of the peculiar kinematics of the 
extra-planar gas (negative rotational gradients and inflow)
which cannot be reproduced in purely 
galactic fountain models \cite{col02,hea06,fb06}.
Fraternali \& Binney (2008) \cite{fb08} have explored this possibility by 
including accretion of low angular momentum gas from the IGM in their models, 
and have calculated the amount of accreting material necessary to reproduce the observations of NGC\,891 and NGC\,2403.
They have found that the accretion rates needed are very close to the 
respective SFRs. 

In conclusion, the extra-planar gas seems to consist of two parts: a large 
one from galactic fountains and a smaller part accreted from intergalactic 
space. There is direct (HVCs in our galaxy and filaments in external galaxies) 
and indirect (rotational velocity gradients) evidence for the accretion from 
outside.  
Accretion rates range from a minimum of about $0.1-0.2 \moyr$ to values possibly 
ten times higher, as needed to meet SFR requirements.
It is clear that the observed extra-planar gas cannot be purely and 
totally made up from accretion: on the one hand, accretion rates would 
be unrealistically high 
($\sim 30 \moyr$ for NGC\,891), and, on the other, there is strong observational
evidence that galactic fountains do 
take place and have a major part in building up the halo (see \ref{halos}).

\section{Extended, warped \hi\ outer disks}
\label{warps}

\begin{figure*}
\centering
\includegraphics[width=0.85\textwidth]{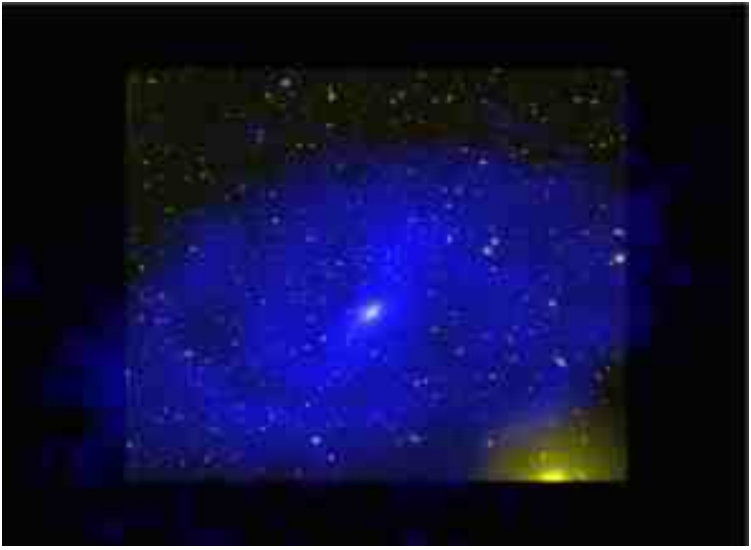}
\caption{Composite optical (yellow) and \hi\ (blue) image of the dwarf galaxy 
NGC\,2915 from Meurer et al.\ (1996) \cite{meur96}.}
\label{f_n2915}
\end{figure*}

The \hi\ disks of the large majority of spiral galaxies are known to extend
considerably beyond the bright optical disk.  
Broeils and Rhee (1997) \cite{broeils97} find for the ratio of the \hi\ 
radius (defined at a surface density level of $1 \mopc$, 
$N_{\rm H}=1.3 \times 10^{20} \cm$)
to the optical radius (R$_{25}$) an average value R$_{\hi}/$R$_{25}=1.7\pm$0.5.
In some cases the \hi\ extends much further out, to several optical radii. 
Striking examples are DDO\,154 \cite{krum84},
NGC\,4449 \cite{baj94},
NGC\,2915 \cite{meur96},
and NGC\,3741 \cite{begum05, gent07}.
This large extent of the \hi\ disks has made it possible to trace the rotation 
curves far beyond the bright stellar disk and thus provide the crucial evidence 
for the existence of dark matter in spiral galaxies. 
We consider the possibility that these outer layers have accumulated 
from tidal debris or infall of gas clouds, as described in section 2, and that 
they now form a reservoir of fresh gas for fuelling star formation 
in the inner regions.  

\begin{figure*}
\centering
\includegraphics[width=\textwidth]{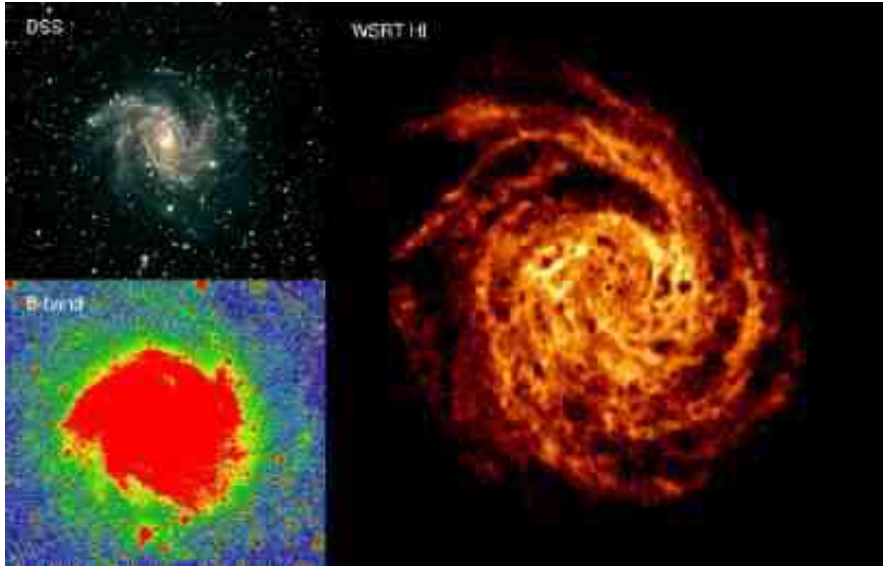}
\caption{Comparison between optical images and total \hi\ map for NGC\,6946.
All images are on the same scale.
Top left: colour composite of the Digitized Sky Survey plates. 
Bottom left: deep B-band image from Ferguson et al.\ (1998) \cite{ferg98}.
Right: deep WSRT total \hi\ map from Boomsma (2007) \cite{boo07}. 
Column densities range from $6 \times 10^{19}\cm$ to $3 \times10^{21}\cm$.}
\label{f_n6946}
\end{figure*}

\begin{figure*}
\centering
\includegraphics[width=\textwidth]{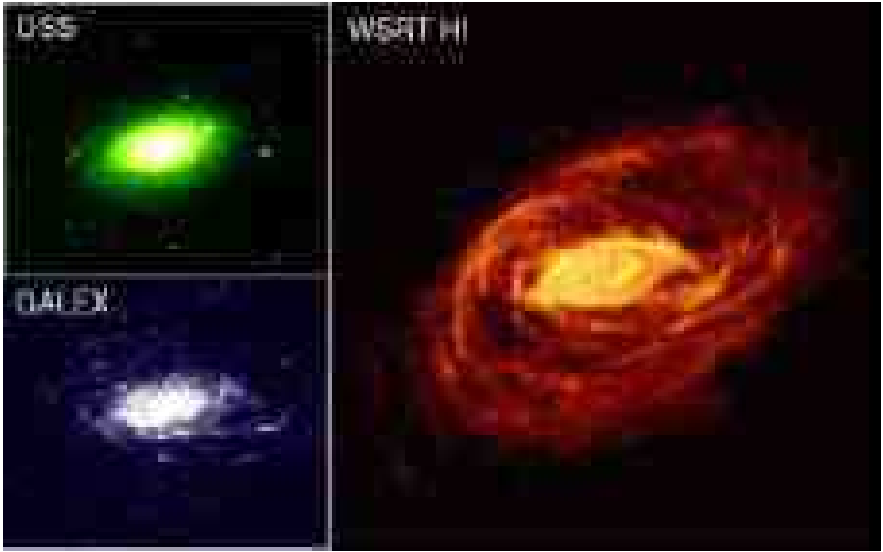}
\caption{Optical (DSS), GALEX and WSRT \hi\ images of the warped galaxy NGC\,5055 
(all on the same scale).  Column densities range from about $3 \times 10^{19}\cm$ to $1 \times 10^{21}\cm$ \cite{bat06}.}
\label{f_n5055}
\end{figure*}

Here, we show the \hi\ images of the blue compact dwarf galaxy NGC\,2915 
(Fig.\ \ref{f_n2915}, \cite{meur96})
and of the two spiral galaxies
NGC\,6946 (Fig.\ \ref{f_n6946}, \cite{boo07}) and
 NGC\,5055 (Fig.\ \ref{f_n5055}, \cite{bat06}).
Generally, in the galaxies studied with sufficient resolution,
the outer \hi\ layers show spiral features (see e.g. NGC\,5055). 
Strong spiral arms are seen in the outer parts of NGC\,6946; 
well-developed arms are also observed in the low luminosity galaxies 
NGC\,2915 and NGC\,3741 \cite{gent07}. 
It is remarkable that there is such pronounced spiral structure in the 
outer regions of spirals where dark matter dominates and even in the dwarfs 
where the dark halo is believed to be predominant everywhere. 
The question is, therefore, 
whether these systems have light disks surrounded by massive dark halos 
or, rather, have heavy and dark disks.

\begin{figure*}
\centering
\includegraphics[width=\textwidth]{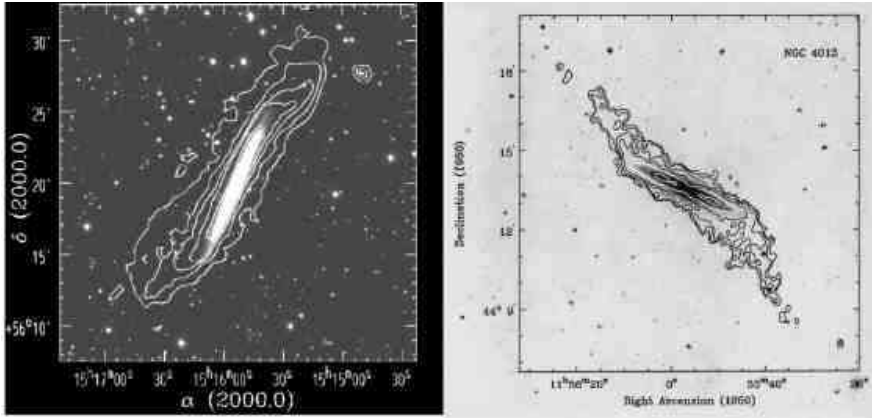}
\caption{Total \hi\ maps (contours) of two warped edge-on 
galaxies overlaid on optical images.
Left: NGC\,5907 from Shang et al. \cite{shang98}. Right: 
NGC\,4013 from Bottema et al. \cite{bot96}.}
\label{f_warps}
\end{figure*}

These gaseous outer parts of disks are usually warped. This is especially  
clear in edge-on galaxies.
In systems viewed at lower inclination angles, the warping 
is inferred from the velocity field. 
Prominent examples of warps are those of the edge-on galaxies 
NGC\,5907 \cite{san76,shang98} and NGC\,4013 \cite{bot96}, 
shown in Fig.\ \ref{f_warps}, and of the less inclined galaxy NGC\,5055 \cite{bat06} shown in Fig.\ \ref{f_n5055}.
These are also among the most symmetrical. 
In general, warps are quite asymmetric, like the warp of our galaxy.
Warps seem to be quite common: according to Bosma (1991) \cite{bosma91} the 
fraction of warped \hi\ disks is at least 50\%. 
A recent \hi\ study of 26 edge-on galaxies \cite{ruiz02}
has shown that 20 galaxies are warped and that all galaxies with \hi\ 
extending beyond the stellar disk are warped to some degree.
There are also edge-on galaxies, such as NGC\,4565, which clearly show the 
presence of an optical warp.
As a result of a statistical study of 540 edge-on galaxies, Reshetnikov \& 
Combes (1999) \cite{resh99} present a sample of 60 of the clearest and strongest 
among the 174 S-shape optical warps found. 
The existence of stellar warps is not surprising.
In a galaxy seen less inclined, such as NGC\,5055, it is possible to see from its 
GALEX image (Fig.\ \ref{f_n5055}, see also \cite{thi07}) that stars have formed 
in the warped gaseous layer, in correspondence with the \hi\ spiral features.
The systematic properties of warps have
been investigated by Briggs (1990) \cite{bri90} and 
more recently by Jozsa (2007) \cite{jozsa07}. 

Although various possibilities for the origin and persistence of warps
have been discussed, there still is no satisfactory
explanation. The possibility that warps are the consequence of accretion of gas 
with a slewed angular momentum due to cosmic infall has been suggested by  
Ostriker \& Binney (1989) \cite{ostr89} and  by Jiang \& Binney (1999)
\cite{jiang99}.

The amount of \hi\ located in the outer parts of spiral galaxies, beyond R$_{25}$,
is on average approximately equal to the amount of \hi\ in the bright inner 
stellar disk \cite{broeils97}. 
It ranges, therefore, from 10$^8$ to 10$^{10} \mo$ from the small to the 
large galaxies. 
Potentially, this gas represents a huge reservoir available for 
the replenishment of the inner disk where the star-formation rate is higher.
As far as we know, however, there is no direct evidence for a radial inflow of 
this cold gas toward the centre. 
In the case of NGC 2403 such a radial inflow, of about 15$\kms$ \cite{fra02a}, 
has been found for the halo gas, but not for the disk.
The \hi\ velocity fields would reveal the presence of large-scale 
inflows larger than 5 $\kms$.
The signature is the non-orthogonality of major and minor axis.
To our knowledge, no such inflow motions have been reported. In order to flow inward, the outer gas has to loose part of its angular momentum. It is not clear how this can happen and, therefore, whether such a process of inflow would work.
An overall radial motion is probably unrealistic. 
However, often the outer \hi\ is not in circular orbits: bars, oval distortions and lopsided structures (see Section \ref{lopsidedness}) are very common and 
may play a role.
As an example, to reach an infall rate of $1 \moyr$ one would need to move 
about $2 \times 10^9 \mo$ of gas from the outer to the inner parts with a 
radial inward speed of 5 $\kms$.
This would take about $2 \times 10^9$ yr.

In conclusion, the possibility envisaged here is that gas accreted from 
satellites or directly from the IGM is deposited in the outermost parts of 
galactic disks. These outer layers of gas and, to some extent, also stars 
are characterized by spiral features and large-scale warping.
They form a reservoir of gas which in turn could, from there, slowly migrate to 
replenish the inner parts of galaxies.

\section{Lopsidedness}
\label{lopsidedness}

\begin{figure*}
\centering
\includegraphics[width=\textwidth]{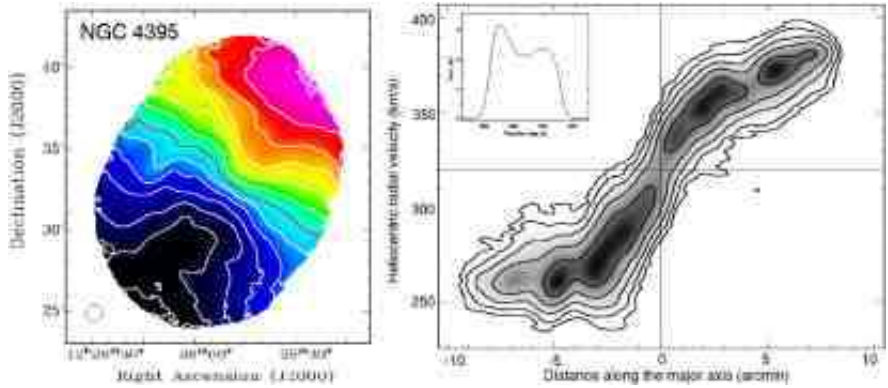}
\caption{Left: velocity field of NGC\,4395 at 1$'$ resolution
(blue is approaching). The contours are at 10 $\kms$ intervals and 
range from 260 $\kms$ (SE) to 380 $\kms$ (NW).
Right: position-velocity plot along the major axis of NGC\,4395
and global \hi\ profile (Heald \& Oosterloo 2008, in preparation).
}
\label{f_n4395}
\end{figure*}

Infall of substantial amounts of gas and stars may have observable effects 
on the disks of spiral galaxies. One of these is the lopsidedness 
in the \hi\ density distribution and in the disk kinematics, which has been known for many years and seems to be a common phenomenon among spirals.  
Since the first study \cite{bal80}, based on a
small number of objects, much new evidence has become available. The
frequency of asymmetries among spiral galaxies has been estimated from
the global \hi\ profiles of a large sample of field
galaxies \cite{rich94}. Examples of asymmetric global \hi\ profiles are shown 
in Fig.\ \ref{f_m101} for M\,101 and in Fig.\ \ref{f_n4395}  for NGC\,4395.
About 20\% of the systems examined
showed strong asymmetries and up to more than 50\% of the whole 
sample showed some mild asymmetries.  This result has been
confirmed by a 21-cm \hi\ survey of 104 isolated galaxies with the
Green Bank 43-m telescope \cite{hay98} and also by more recent observations which image the \hi\ distribution and the kinematics of a large sample of galaxies  (WHISP). At least one half of about 300 objects from WHISP shows  
some  lopsidedness either in the \hi\ distribution or in the kinematics or in both. Deviations from axial symmetry seem, therefore, to be the rule 
rather than the exception. 
It should be noted that, in general, these lopsided galaxies are not
interacting systems and that, therefore, the lopsidedness cannot be explained 
as a present tidal effect.

The lopsidedness is generally  present in the \hi\ density distribution 
as in M\,101 (see Fig.\ \ref{f_m101})
and also in the kinematics. 
In some cases, such as NGC\,4395 (see Fig.\ \ref{f_n4395}), 
the asymmetry is only seen in the kinematics:
on one side of the galaxy the rotation curve rises more slowly 
(almost solid body) and reaches 
the flat part at larger radii than on the other side. 
This is the signature of the
kinematic lopsidedness as discussed by Swaters et al.\ (1999) \cite{swa99}. 
In such cases there seems to be a well-defined pattern that pervades the whole velocity field and may be related to a lopsided potential.

\begin{figure*}
\centering
\includegraphics[width=\textwidth]{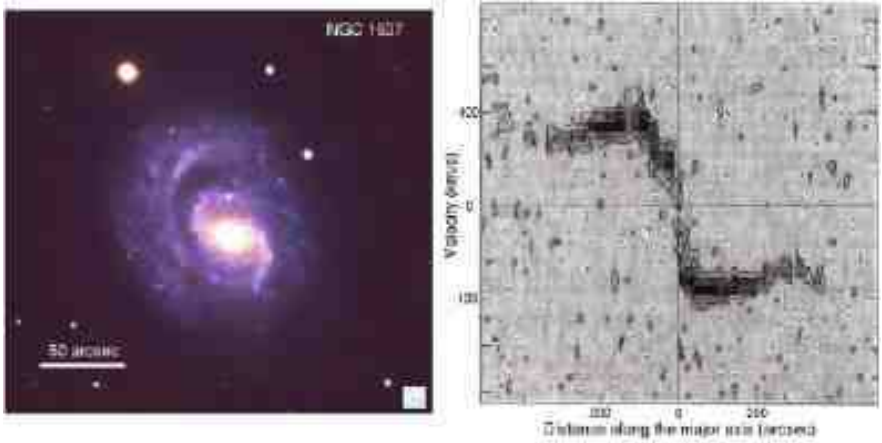}
\caption{Left: optical image of NGC\,1637. 
Right: position-velocity plot along the major axis of NGC\,1637 \cite{rob01},
North is on the left.
}
\label{f_n1637}
\end{figure*}

Although the morphological lopsidedness is most clearly seen in the
\hi\ data, there is often evidence of asymmetries
also in the distribution of light, in the B-band as well as in the I and K
bands \cite{rix95, zar97, korn98}.
One of the most striking cases is that
of the spiral galaxy NGC\,1637 (Fig.\ \ref{f_n1637}), which is exceptionally
lopsided in blue light as well as in the near infrared \cite{block94},
indicating that the lopsidedness is present not only in the
young stellar population but also in the old stellar disk.
In this galaxy there is clear evidence for a kinematical 
anomaly  associated with the morphological asymmetry.
Indeed, \hi\ observations with the VLA \cite{rob01}
reveal a peculiarity in the velocity structure in correspondence with 
the northern anomalous arm. This shows up as a step in the position-velocity 
diagram along the major axis (at about $-60''$ in the p-v diagram in 
Fig.\ \ref{f_n1637}) implying a local deviation of at least $50 \kms$ 
from circular motion. 
This peculiar feature closely resembles the strong kinematic anomaly 
and the strong \hi\ arm in the southern part of M 101 \cite{kam93}. 
This suggests that the lopsidedness of NGC\,1637 is not only manifest in the optical appearance, but it is also present in the disk dynamics. 
As Block et al.\  \cite{block94} point out, 
this is either an extreme m$=$1 asymmetry, and in such case its persistence 
should be explained, or it is the effect of a recent tidal interaction. 
We note that no companion for NGC\,1637 is found on the Palomar Observatory 
Sky Survey within one degree.

All these facts suggest that the phenomenon of lopsidedness in spiral 
galaxies is quite common and structural for the disk.
It is not clear what its origin is and how persistent it is. 
Minor mergers \cite{zar97} and tidal interactions \cite{korn02}
have been suggested as possible causes. 
Bournaud et al.\ (2005) \cite{bou05}, however, 
have found from numerical simulations  
that galaxy interactions and mergers are not sufficient to explain all the 
properties of the observed lopsided galaxies and have concluded that cosmological 
accretion of gas on galactic disks must be responsible. 
At any rate, it seems likely that many of the asymmetries, especially 
some of those revealed by the \hi\ distribution and kinematics, as for 
instance in M\,101, are transient phenomena and are due to recent accretion 
events such as those described above in Section \ref{interactions}.

\section{Intergalactic \hi\ clouds}
\label{IGM}

In the previous sections we have discussed direct and indirect evidence 
on the accretion of cold gas in galaxies.
We have argued that part of this accretion may come from the
merging with gas-rich dwarf companions and part may come {\it directly}
from the IGM in the form of gas clouds or filaments 
without stellar counterpart. 
It is natural, therefore, to ask what is the role of 
dwarfs in bringing in the gas and whether \hi\ clouds exist in 
intergalactic space, away from galaxies.

Several observations of the fields around galaxies and of groups
similar to the Local Group have been undertaken \cite{zwa01,blok02,pisano04,pisano07}
as well as large blind surveys, such as HIPASS \cite{barn01} and the ongoing
large survey ALFALFA (e.g.\ \cite{gio05}).
Most of these observations have a sensitivity to detect clouds of \hi\ with a
mass of about 10$^7 \mo$ (assuming a profile width of 30 $\kms$).
One survey \cite{kovac07} has the sensitivity (over large volumes of space) to detect smaller objects, down to masses of 10$^6 \mo$.

The results can be summarized as follows:

i) There is no evidence for a significant population of isolated \hi\ clouds
(so-called dark galaxies) in the IGM.
Kova\v{c} (2007) finds no clouds without optical counterpart down to 10$^6 \mo$.  
The first results from ALFALFA show that only 3\% of their \hi\ sources
are not detected in the optical \cite{hay07, gio07}.
There is a hint in Kova\v{c}'s survey, that galaxies with low \hi\
masses are missing in the Local Volume.
This could indicate that there is a lower limit
to the gas mass of field galaxies of around a few times 10$^6 \mo$.

ii) In galaxy groups no clouds are detected, with detection limits of
10$^7 \mo$.
By analogy, this is an argument against the hypothesis that the HVCs
are members of the Local Group with masses above 10$^7 \mo$ \cite{bli99},
in line with the recent distance determinations \cite{wak07, wak08} and
with the conclusion that HVCs are located in the Galactic halo
(see Section \ref{MW}) and have masses lower than 10$^7 \mo$.

These results also agree with what is known from the very deep \hi\ 
observations of some nearby galaxies, like those of NGC\,891 and NGC\,6946 reported above.
Although the areas surveyed around these galaxies are not very extended, 
as they usually correspond to only a few times the galactic \hi\ images,
they do seem to be empty and, remarkably, all the clouds detected 
(with masses of 10$^6-10^7 \mo$) are located very close to or within the halo regions of the galaxies. Yet, some of these clouds are likely to have an  intergalactic origin and not to be the result of galactic fountains.

To sum up, there is no evidence for the presence of a population 
of gas clouds in intergalactic space capable of accounting for the gas 
now observed near galaxies (halos and surroundings) and, above all, 
capable of fuelling the whole process of star formation.

Could all the gas needed for star formation have been brought in by 
dwarf companions?
The \hi\ mass function shows that most of the \hi\ in the local universe is in
large galaxies, with masses above about $10^9 \mo$ \cite{zwa05a}.
If we assume that large galaxies accrete gas only via minor mergers with
smaller galaxies and we require that the rate of gas accretion is
about $1 \moyr$ ($\sim$SFR), then all small 
galaxies would have been accreted by the large galaxies in a timescale of the 
order of $1 \times 10^9$ yr.
Clearly, an accretion rate of $1 \moyr$ purely in the form of small
galaxies cannot be sustained for much longer than one Gyr.
This is inconsistent with the constancy of star formation in the
Milky Way and also with the results of absorption studies of Damped
Lyman Alpha (DLA) systems \cite{proch05}, which show that 
the cosmic integral \hi\ mass density and the 
column density distribution of galaxies have evolved very
little in the last 10 Gyr \cite{zwa05b}.

It is clear from the above considerations that constant gas accretion 
rates as high as required for star formation cannot be reached via mergers 
with satellite galaxies and that, therefore, a substantial amount of gas
accretion  must come  {\it directly} from the IGM.
Most of this accretion must take the form of gas clouds (or filaments).
The reason why such clouds are not observed in intergalactic space
remains an open question.
Possibilities are that:
i) the clouds are confined near the galaxies by the hot galactic
halos, whereas they disperse in the IGM;
ii) the clouds are intrinsically very small
(smaller that $\approx 10^6 \mo$) and the accretion takes the form of
a continuous drizzle of gas. Note that only the \hi\ masses have to be small 
and a certain amount of dark matter cannot be excluded.
iii) \hi\ clouds are produced by large-scale cooling flows
of the IGM in the regions surrounding the galaxies \cite{kau06, peek07}.

\section{Conclusion}
\label{conclusions}

We have drawn attention to a number of  results from \hi\ observations 
of nearby spiral galaxies which bear directly or indirectly on cold 
gas accretion. 
There are large \hi\ complexes, with and without stellar counterparts, 
in the neighborhood of galaxies and in close interaction with them. 
There is little doubt about their extragalactic origin. 
Moreover, many galaxies have extra-planar gas components.
Although most of this extra-planar gas must come from galactic 
fountains, a fraction of it, being in filaments and massive clouds like the
Galactic HVCs, is likely to have an external origin.
From the study of the gas seen in the neighborhood of galaxies and 
in their halos (the fountain gas excepted) we estimate a ``visible'' accretion 
rate of at least $0.2 \moyr$. 

Furthermore, there are large amounts of \hi, 
in the mass range $10^{8-10} \mo$, in the warped outer  
galactic disks which could be a gas reservoir for replenishing the inner parts and fuelling star formation. 
In some cases (e.g.\ NGC\,5055) it is obvious that this gas, because of its 
symmetrical and regular structure and long timescales, must have been there 
for a long time (several Gyr). 
It is conceivable, however, that in other cases 
the outer gas layers have been accumulated from outside more recently 
and that warping and spiral arms are related to such a process.
Finally, there is the phenomenon of disk lopsidedness, which is  poorly 
understood, but may be pointing to recent infall.
Although both the warped outer layers and the lopsidedness may well be the 
effect of accretion from intergalactic space, estimates of infall rates are
difficult to obtain.

The visible gas accretion rate, estimated above, does not seem to be 
sufficient to account for the star formation in galaxies 
($\approx 1 \moyr$). Direct, indisputable evidence for the required levels of cold 
gas accretion does not exist.
There is, however, the puzzle of the peculiar kinematics of the extra-planar 
gas (overall negative vertical gradients in the rotational velocities and inflow motion) that could be solved by invoking infall of gas from outside carrying low angular momentum.
It seems that this would give accretion rates of the order of the SFRs.

Finally, the question arises of where the infalling gas, we hypothesized
above, could originate from.
Gas-rich dwarfs seem only to be able to account for a small fraction of the
required gas
and thus most of the accretion should come directly from the IGM.
There is, however, no evidence for a population of large \hi\ clouds
in intergalactic space, in regions away from galaxies. 
Therefore, how and in what form gas infall onto galaxies takes place
remains an open question and a challenge for future investigations.

\begin{acknowledgements} 
We thank Rense Boomsma for providing the NGC 6946 images and George Heald for the NGC\,4395 figure. 

We are grateful to James Binney, Raffaella Morganti, Eline Tolstoy, Monica Tosi 
and Martin Zwaan for helpful comments and stimulating discussions.
\end{acknowledgements}

%REFERENCES


\begin{thebibliography}{}

\bibitem[Bajaja, Huchtmeier \& Klein]{baj94} Bajaja E., Huchtmeier W.K., Klein U. (1994), A\&A, 285, 385

\bibitem[Balcells \& Sancisi(1996)]{bal96} Balcells M.\ \& Sancisi R. (1996), AJ, 111, 1053

\bibitem[Baldwin, Lynden-Bell \& Sancisi(1980)]{bal80} Baldwin J.E., Lynden-Bell D., Sancisi R. (1980), MNRAS, 193, 313

\bibitem[Barbieri et al.(2005)]{barb05} Barbieri C.V., Fraternali F., Oosterloo T., Bertin G., Boomsma R., Sancisi R. (2005), A\&A, 439, 947

\bibitem[Barnab\`e et al.(2006)]{barn06} Barnab\`e M., Ciotti L., Fraternali 
F., Sancisi R., 2006, A\&A, 446, 61

\bibitem[Barnes et al.(2001)]{barn01} Barnes D.G., Staveley-Smith L., de Blok W.J.G., Oosterloo T., Stewart I.M., Wright A.E., Banks G.D., Bhathal R., Boyce P.J., Calabretta M.R., Disney M.J., Drinkwater M.J., Ekers R.D., Freeman K.C., Gibson B.K., Green A.J., Haynes R.F., te Lintel Hekkert P., Henning P.A., Jerjen H., Juraszek S., Kesteven M.J., Kilborn V.A., Knezek P.M., Koribalski B., Kraan-Korteweg R.C., Malin D.F., Marquarding M., Minchin R.F., Mould J.R., Price R.M., Putman M.E., Ryder S.D., Sadler E.M., Schröder A., Stootman F., Webster R.L., Wilson W.E., Ye T. (2001), MNRAS, 322, 486

\bibitem[Battaglia et al.(2005)]{bat06} Battaglia G., Fraternali F., Oosterloo T., Sancisi R. (2005), A\&A, 447, 49

\bibitem[Begum, Chengalur \& Karachentsev(2005)]{begum05} Begum A., Chengalur J.N., Karachentsev I.D. (2005), A\&A, 433, L1

\bibitem[Belokurov et al.(2007)]{belo07} Belokurov V., Evans N.W., Irwin M.J., Lynden-Bell D., Yanny B., Vidrih S., Gilmore G., Seabroke G., Zucker D.B., Wilkinson M.I., Hewett P.C., Bramich D.M., Fellhauer M., Newberg H.J., Wyse R.F.G., Beers T.C., Bell E.F., Barentine J.C., Brinkmann J., Cole N., Pan K., York D.G. (2007), ApJ, 658, 337

\bibitem[Binney, Dehnen \& Bertelli(2000)]{BinneyDB} Binney J.J., Dehnen W., Bertelli G. (2000), MNRAS, 318, 658

\bibitem[Blitz et al.(1999)]{bli99} Blitz L., Spergel D.N., Teuben P.J., Hartmann D., Burton W.B. (1999), ApJ, 514, 818

\bibitem[Block et al.(1994)]{block94} Block D.L., Bertin G., Stockton A., Grosbol P., Moorwood A.F.M., Peletier R.F. (1994), A\&A, 288, 365

\bibitem[Bond et al.(1991)]{Bond} Bond J.R., Cole S., Efstathiou G., Kaiser N. (1991), ApJ, 379, 440

\bibitem[Boomsma et al.(2005a)]{boo05a} Boomsma R., Oosterloo T.A., Fraternali F., van der Hulst J.M., Sancisi R. (2005a), A\&A, 431, 65

\bibitem[Boomsma et al.(2005b)]{boo05b} Boomsma R.,  Oosterloo T.,  Fraternali F.,  van der Hulst J.M.,  Sancisi R. (2005b), in: ed.\ R.\ Braun, {\it Extra-planar Gas}, Dwingeloo, ASP Conf.\ Series, Vol.\ 431, p.\ 65

\bibitem[Boomsma(2007)]{boo07} Boomsma R. (2007), {\it PhD Thesis}, University of Groningen

\bibitem[Bosma(1991)]{bosma91} Bosma A. (1991), in: eds. S. Casertano, P. Sackett \& F. Briggs, {\it Warped disks and inclined rings around galaxies}, Cambridge University Press, Cambridge, p. 181

\bibitem[Bottema(1996)]{bot96} Bottema R. (1996), A\&A, 306, 345

\bibitem[Boulanger \& Viallefond(1992)]{boul92} Boulanger F.\ \& Viallefond F. (1992), A\&A, 266, 37

\bibitem[Bournaud et al.(2005)]{bou05} Bournaud F., Combes F., Jog, C.J., Puerari I. (2005), A\&A, 438, 507

\bibitem[Braun et al.(1994)]{braun94} Braun R., Walterbos R.A.M., Kennicutt R.C., Tacconi L.J. (1994), ApJ, 420, 558

\bibitem[Bregman(1980)]{breg80} Bregman J.N. (1980), ApJ, 236, 577

\bibitem[Briggs(1990)]{bri90} Briggs F.H. (1990), ApJ, 352, 15

\bibitem[Broeils \& Rhee(1997)]{broeils97} Broeils A.H.\ \& Rhee M.-H. (1997), A\&A, 324, 877

\bibitem[Bruns et al.(2005)]{bruns05} Br\"uns C., Kerp J., Staveley-Smith L., Mebold U., Putman M.E., Haynes R.F., Kalberla P.M.W., Muller E., Filipovic M.D. (2005), A\&A, 432, 45

\bibitem[Carilli \& Van Gorkom(1992)]{car92} Carilli C.L.\ \& Van Gorkom J.H. (1992), ApJ, 399, 373

\bibitem[Chaves \& Irwin(2001)]{cha01} Chaves T.A.\ \& Irwin J.A. (2001), ApJ, 557, 646

\bibitem[Collins, Benjamin \& Rand(2002)]{col02} Collins J.A., Benjamin R.A., Rand R.J. (2002), ApJ, 578, 98

\bibitem[da Costa et al.(1998)]{dacosta98} da Costa L.N., Willmer C.N.A., Pellegrini P.S., Chaves O.L., Rit\'e C., Maia M.A.G., Geller M.J., Latham D.W., Kurtz M.J., Huchra J.P., Ramella M., Fairall A.P., Smith C., L\'ipari S. (1998), AJ, 116, 1

\bibitem[Dahlem et al.(2005)]{dahlem05} Dahlem M., Ehle M., Ryder S.D., Vlaji\'c M., Haynes R.F. (2005), A\&A, 432, 475

\bibitem[de Blok et al.(2002)]{blok02} de Blok W.J.G., Zwaan M.A., Dijkstra M., Briggs F.H., Freeman K.C. (2002), A\&A, 338, 43

\bibitem[Dekel \& Birnboim(2006)]{dekel06} Dekel A.\ \& Birnboim Y. (2006), MNRAS, 368, 2

\bibitem[Ferguson et al.(1998)]{ferg98} Ferguson A.M.N., Wyse R.F.G., Gallagher J.S., Hunter D.A. (1998), ApJ, 506, 19L

\bibitem[Ferguson et al.(2002)]{ferg02} Ferguson A.M.N., Irwin M.J., Ibata R.A., Lewis G.F., Tanvir N.R. (2002), AJ, 124, 1452 

\bibitem[Fisher \& Tully(1976)]{fish76} Fisher J.R.\ \& Tully R.B. (1976), A\&A, 53, 397

\bibitem[Fraternali et al.(2001)]{fra01} Fraternali F., Oosterloo T., Sancisi R., van Moorsel G. (2001), ApJ, 562, 47

\bibitem[Fraternali et al.(2002a)]{fra02a} Fraternali F., van Moorsel G., Sancisi R., Oosterloo T. (2002a), AJ, 123, 3124

\bibitem[Fraternali et al.(2005)]{fra05} Fraternali F., Oosterloo T., Sancisi R., Swaters R. (2005), in: ed.\ R.\ Braun, {\it Extra-planar Gas}, Dwingeloo, ASP Conf.\ Series, Vol.\ 331, p.\ 239

\bibitem[Fraternali \& Binney(2006)]{fb06} Fraternali F.\ \& Binney J.J. (2006), MNRAS, 366, 449

\bibitem[Fraternali \& Binney(2008)]{fb08} Fraternali F.\ \& Binney J.J. (2008), MNRAS, in press

\bibitem[Fukugita \& Peebles(2004)]{fuku04} Fukugita M.\ \& Peebles P.J.E. (2004), ApJ, 616, 643

\bibitem[Garc\'ia-Ruiz, Sancisi \& Kuijken(2002)]{ruiz02} Garc\'ia-Ruiz I., Sancisi R., Kuijken K. (2002), A\&A, 394, 769

\bibitem[Gentile et al.(2007)]{gent07} Gentile G., Salucci P., Klein U., Granato G.L. (2007), MNRAS, 375, 199

\bibitem[Giovanelli et al.(2005)]{gio05} Giovanelli R., Haynes M.P., Kent B.R., Perillat P., Saintonge A., Brosch N., Catinella B., Hoffman G.L., Stierwalt S., Spekkens K., Lerner M.S., Masters K.L., Momjian E., Rosenberg J.L., Springob C.M., Boselli A., Charmandaris V., Darling J.K., Davies J., Lambas D.G., Gavazzi G., Giovanardi C., Hardy E., Hunt L.K., Iovino A., Karachentsev I.D., Karachentseva V.E., Koopmann R.A., Marinoni C., Minchin R., Muller E., Putman M., Pantoja C., Salzer J.J., Scodeggio M., Skillman E., Solanes J.M., Valotto C., van Driel W., van Zee L. (2005), AJ, 130, 2598

\bibitem[Giovanelli et al.(2007)]{gio07} Giovanelli R., Haynes M.P., Kent B.R., Saintonge A., Stierwalt S., Altaf A., Balonek T., Brosch N., Brown S., Catinella B., Furniss A., Goldstein J., Hoffman G.L., Koopmann R.A., Kornreich D.A., Mahmood B., Martin A.M., Masters K.L., Mitschang A., Momjian E., Nair P.H., Rosenberg J.L., Walsh B. (2007), AJ, 133, 2569

\bibitem[Haynes et al.(1998)]{hay98} Haynes M.P., Hogg D.E., Maddalena R.J., Roberts M.S., Van Zee L. (1998), AJ, 115, 62

\bibitem[Haynes(2007)]{hay07} Haynes M.P. (2007), in: eds. A.\ Bridle, J.\ Condon \& G.\ Hunt, {\it Frontiers of Astrophysics}, ASP Conf.\ Series, 18-21 June 2007, (astro-ph/0708.2547)

\bibitem[Heald et al.(2006)]{hea06} Heald G.H., Rand R.J., Benjamin R.A., Bershady M.A. (2006), ApJ, 647, 1018

\bibitem[Heald \& Oosterloo(2008)]{heald08} Heald G. \& Oosterloo T. (2008), in: eds. J.G. Funes, S.J. \& E.M. Corsini, {\it Formation and Evolution of Galaxy Disks}, Rome, 1-5 October 2007, (astro-ph/0712.1184).

\bibitem[Heckman et al.(1982)]{heck82} Heckman T.M., Sancisi R., Sullivan III W.T., Balick B. (1982), MNRAS, 199, 425

\bibitem[Helmi \& de Zeeuw(2000)]{helmi00} Helmi A.\ \& Tim de Zeeuw T.P. (2000), MNRAS, 319, 657

\bibitem[Helmi \& White(2001)]{helmi01} Helmi A.\ \& White S.D.M. (2001), MNRAS, 323, 529

\bibitem[Hibbard et al.(2001)]{hibba01} Hibbard J.E., van Gorkom J.H., Rupen M.P., Schiminovich D. (2001), in: eds. J.E. Hibbard, J.H. van Gorkom \& M.P. Rupen, {\it Gas and Galaxy Evolution}, ASP Conf.\ Series, Vol.\ 240, p.\ 657

\bibitem[Ibata, Gilmore \& Irwin(1994)]{ibata94} Ibata R.A., Gilmore G., Irwin M.J. (1994), Nature, 370, 194 

\bibitem[Ibata et al.(2001)]{ibata01} Ibata R., Irwin M., Lewis G., Ferguson A.M.N., Tanvir N. (2001), Nature, 412, 49 

\bibitem[Irwin(1994)]{irw94} Irwin J.A. (1994), ApJ, 429, 618

\bibitem[Irwin \& Chaves(2003)]{irw03} Irwin J.A.\ \& Chaves T. (2003), ApJ, 585, 268

\bibitem[Jiang \& Binney(1999)]{jiang99} Jiang I.-G.\ \& Binney J.J. (1999), MNRAS, 303, L7

\bibitem[Jozsa(2007)]{jozsa07} Jozsa G.I.G. (2007), A\&A, 468, 903

\bibitem[Kamphuis \& Briggs(1992)]{kam92} Kamphuis J.\ \& Briggs F. (1992), A\&A, 253, 335

\bibitem[Kamphuis(1993)]{kam93} Kamphuis J.J. (1993), {\it PhD Thesis}, University of Groningen

\bibitem[Kamphuis \& Sancisi(1994)]{kam94} Kamphuis J.\ \& Sancisi R. (1994), in: eds. G. Hensler, C. Theis \& J.S. Gallagher, {\it Panchromatic View of Galaxies.\ Their Evolutionary Puzzle}, Editions Frontiers, p. 317

\bibitem[Kaufmann et al.(2006)]{kau06} Kaufmann T., Mayer L., Wadsley J., Stadel J., Moore B. (2006), MNRAS, 370, 1612

\bibitem[Kere\v{s} et al.(2005)]{keres05} Kere\v{s} D., Katz N., Weinberg D.H., Dav\'e R. (2005), MNRAS, 363, 2

\bibitem[Kornreich et al.(1998)]{korn98} Kornreich D.A., Haynes M.P., Lovelace R.V.E. (1998), AJ, 116, 2154

\bibitem[Kornreich et al.(2002)]{korn02} Kornreich D.A., Lovelace R.V.E., Haynes M.P. (2002), ApJ, 580, 705

\bibitem[Kova\'c(2007)]{kovac07} Kova\v{c} K. (2007), {\it PhD Thesis}, University of Groningen

\bibitem[Kregel \& Sancisi(2001)]{kre01} Kregel M.\ \& Sancisi R. (2001), A\&A, 376, 5

\bibitem[Krumm \& Burstein(1984)]{krum84} Krumm N.\ \& Burstein, D. (1984), AJ 89, 1319

\bibitem[Lacey \& Cole(1993)]{LaceyCole} Lacey C.\ \& Cole S. (1993), MNRAS, 262, 627

\bibitem[Larson(1972)]{lars72} Larson R.B. (1972), Nature, 236, 21L

\bibitem[Larson(1980)]{lars80} Larson R.B.,Tinsley B.M., Caldwell C.N. (1980), ApJ, 237, 692

\bibitem[Lee et al.(2001)]{lee01} Lee S.-W., Irwin J.A., Dettmar R.-J., Cunningham C.T., Golla G., Wang Q.D. (2001), A\&A, 377, 759

\bibitem[Malin \& Hadley(1997)]{mal97} Malin D., Hadley B. (1997), PASA, 14, 52

\bibitem[Mathewson et al.(1974)]{mat74} Mathewson D.S., Cleary M.N., Murray J.D. (1974), ApJ, 190, 291

\bibitem[Matteucci(2003)]{matteucci} Matteucci F. (2003), in: ed.\ A. McWilliam \& M. Rauch, {\it Origin and Evolution of the Elements}, Cambridge University Press, (astro-ph/0306034)

\bibitem[Matthews \& Wood(2003)]{mat03} Matthews L.D.\ \& Wood K. (2003), ApJ, 593, 721

\bibitem[Matthews \& Uson(2008)]{mat08} Matthews L.D.\ \& Uson J.M. (2008), AJ, 135, 291

\bibitem[McConnachie et al.(2003)]{mccon03} McConnachie A.W., Irwin M.J., Ibata R.A., Ferguson A.M.N., Lewis G.F., Tanvir N. (2003), MNRAS, 343, 1335 

\bibitem[McNamara et al.(1994)]{mcna94} McNamara B.R., Sancisi R., Henning P.A., Junor W. (1994), AJ, 108, 844

\bibitem[Meurer et al.(1996)]{meur96} Meurer G.R., Carignan C., Beaulieu S.F., Freeman K.C. (1996), AJ 111, 1551

\bibitem[Morganti et al.(2006)]{morganti06} Morganti R., de Zeeuw P.T., Oosterloo T.A., McDermid R.M., Krajnovi\'c D., Cappellari M., Kenn F., Weijmans A., Sarzi M. (2006), MNRAS, 371, 157

\bibitem[Mulder, Van Driel \& Braine(1995)]{mul95} Mulder P.S., Van Driel W., Braine J. (1995), A\&A, 300, 687

\bibitem[Oosterloo(2004)]{oos04} Oosterloo T. (2004), in: eds. H.\ van Woerden, B.P.\ Wakker, U.J.\ Schwarz \& K.S.\ de Boer, {\it High-Velocity Clouds}, Astrophysics and Space Science Library, vol.\ 312, p.\ 125

\bibitem[Oosterloo, Fraternali \& Sancisi(2007)]{oos07} Oosterloo T., Fraternali F., Sancisi R. (2007), AJ, 134, 1019

\bibitem[Oosterloo et al.(2007)]{oos07b} Oosterloo T.A., Morganti R., Sadler E.M., van der Hulst T., Serra P. (2007), A\&A, 465, 787

\bibitem[Ostriker \& Binney(1989)]{ostr89} Ostriker E.C.\ \& Binney J.J. (1989), MNRAS, 237, 785

\bibitem[Peek, Putman \& Sommer-Larsen(2007)]{peek07} Peek J.E.G., Putman M.E., Sommer-Larsen J. (2007), ApJ, submitted, (astro-ph/0705.0357)

\bibitem[Phookun et al.(1992)]{phoo92} Phookun B., Mundy L.G., Teuben P.J., Wainscoat R.J. (1992), ApJ, 400, 516

\bibitem[Pisano, Wilcots \& Elmergreen(1998)]{pisano98} Pisano D.J., Wilcots E.M., Elmergreen B.G. (1998), AJ, 115, 975

\bibitem[Pisano, Wilcots \& Elmergreen(2000)]{pisano00} Pisano D.J., Wilcots E.M., Elmergreen B.G. (2000), AJ, 120, 763

\bibitem[Pisano et al.(2004)]{pisano04} Pisano D.J., Barnes D.G., Gibson B.K., Staveley-Smith L., Freeman K.C., Kilborn V.A. (2004), ApJ, 610, L17

\bibitem[Pisano et al.(2007)]{pisano07} Pisano D.J., Barnes D.G., Gibson B.K., Staveley-Smith L., Freeman K.C., Kilborn V.A. (2007), ApJ, 662, 959

\bibitem[Prochaska, Herbert-Fort \& Wolfe(2005)]{proch05} Prochaska J.X., Herbert-Fort S., Wolfe A.M. (2005), ApJ, 635, 123

\bibitem[Rand(1994)]{rand94} Rand R.J. (1994), A\&A, 285, 83

\bibitem[Reshetnikov \& Combes(1999)]{resh99} Reshetnikov V.\ \& Combes F. (1999), A\&AS 138, 101 

\bibitem[Richter \& Sancisi(1994)]{rich94} Richter O.-G.\ \& Sancisi R. (1994), A\&A, 290, L9

\bibitem[Rix \& Zaritsky(1995)]{rix95} Rix H.-W.\ \& Zaritsky D. (1995), ApJ, 447, 82

\bibitem[Roberts, Hogg \& Schulman(2001)]{rob01} Roberts M., Hogg D.E., Schulman E. (2001), in: eds. J.E. Hibbard, M. Rupen \& J.H. van Gorkom, {\it Gas and Galaxy Evolution}, ASP Conf.\ Series, Vol.\ 240, p.\ 294

\bibitem[Rogstad(1973)]{rog73} Rogstad D.H., Shostak G.S., Rots A.H. (1973), A\&A, 22, 111

\bibitem[Rots et al.(1990)]{rots90} Rots A.H., Bosma A., van der Hulst J.M., Athanassoula E., Crane P.C. (1990), AJ, 100, 387

\bibitem[Rupen(1991)]{rup91} Rupen M.P. (1991), AJ, 102, 48

\bibitem[Sadler, Oosterloo \& Morganti(2001)]{sadler01} Sadler E. M., Oosterloo T., Morganti R. (2001), in: eds. J.G. Funes \& E.M. Corsini, {\it Galaxy Disks and Disk Galaxies}, ASP Conf.\ Series, Vol.\ 230, p.\ 285

\bibitem[Sadler, Oosterloo \& Morganti(2002)]{sadler02} Sadler E. M., Oosterloo T., Morganti R. (2002), in: eds. G.S. Da Costa \& H. Jerjen, {\it The Dynamics, Structure \& History of Galaxies}, ASP Conf.\ Series, Vol.\ 273, p.\ 215

\bibitem[Saglia \& Sancisi(1988)]{sag88} Saglia R.P.\ \& Sancisi R. (1988), A\&A, 203, 28.

\bibitem[Saha, Combers \& Jog(2007)]{saha07} Saha K., Combes F., Jog C.J. (2007), MNRAS, 382, 419

\bibitem[Sancisi(1976)]{san76} Sancisi R. (1976), A\&A, 53, 159 

\bibitem[Sancisi \& Allen(1979)]{san79} Sancisi R., \& Allen R.J. (1979), A\&A, 74, 73

\bibitem[Sancisi et al.(1984)]{san84} Sancisi R., Van Woerden H., Davies R.D., Hart L. (1984), MNRAS, 210, 497

\bibitem[Sancisi(1992)]{san92} Sancisi R. (1992), in: eds.\ by Trinh Xuan Thuan, Chantal Balkowski \& J.\ Tran Thanh Van, {\it Physics of Nearby Galaxies: Nature or Nurture?}, Proceedings of the 27th Rencontre de Moriond, Les Arcs, France, March 15-22, Gif-sur-Yvette: Editions Frontieres, p.\ 31

\bibitem[Sancisi(1999a)]{san99a} Sancisi R. (1999a), in: eds. J.E. Barnes \& D.B. Sanders, IAU Symp.\ 186: {\it Galaxy Interactions at Low and High Redshift}, p.\ 71

\bibitem[Sancisi(1999b)]{san99b} Sancisi R. (1999b), Ap\&SS, 269, 59  

\bibitem[Schaap(2000)]{scha00} Schaap W.E., Sancisi R., Swaters R.A. (2000), A\&A, 356, 49L

\bibitem[Schiminovich et al.(1994)]{schi94} Schiminovich D., Van Gorkom J.H., Van der Hulst J.M., Kasov S. (1994), ApJ, 423, L101 

\bibitem[Schiminovich et al.(1995)]{schi95} Schiminovich D., Van Gorkom J.H., Van der Hulst J.M., Malin D.F. (1995), ApJ, 444, L77

\bibitem[Schiminovich et al.(1997)]{schi97} Schiminovich D., van Gorkom J., van der Hulst T., Oosterloo T., Wilkinson A. (1997), {\it The Nature of Elliptical Galaxies}, 2nd Stromlo Symposium, 116, 362 

\bibitem[Shang et al.(1998)]{shang98} Shang Z., Brinks E., Zheng Z., Chen J., Burstein D., Su H., Byun Y.-I., Deng L., Deng Z., Fan X., Jiang Z., Li Y., Lin W., Ma F., Sun W.-H., Wills B., Windhorst R.A., Wu H., Xia X., Xu W., Xue S., Yan H., Zhou X., Zhu J., Zou Z. (1998), ApJ, 504, 23L

\bibitem[Shapiro \& Field(1976)]{sha76} Shapiro P.R.\ \& Field G.B. (1976), ApJ, 205, 762

\bibitem[Shostak et al.(1982)]{sho82} Shostak G.S., Hummel E., Shaver P.A., Van der Hulst J.M., Van der Kruit P.C. (1982), A\&A, 115, 293

\bibitem[Simkin et al.(1987)]{sim87} Simkin S.M., Van Gorkom J.H., Hibbard J.E., Su H-J. (1987), Science, 235, 1367

\bibitem[Sommer-Larsen(2006)]{som06} Sommer-Larsen J. (2006), ApJ, 644L, 1

\bibitem[Smith(1994)]{smi94} Smith B.J. (1994), AJ, 107, 1695

\bibitem[Swaters, Sancisi \& van der Hulst(1997)]{swa97} Swaters R.A., Sancisi R., van der Hulst J.M. (1997), ApJ, 491, 140 

\bibitem[Swaters et al.(1999)]{swa99} Swaters R.A., Schoenmakers R.H.M., Sancisi R., van Albada T.S. (1999), MNRAS, 304, 330

\bibitem[Tacconi(1986)]{tac86} Tacconi L.J.\ \& Young J.S. (1986), ApJ, 308, 600 

\bibitem[Taramopoulos, Payne \& Briggs(2001)]{tara01} Taramopoulos A., Payne H., Briggs F.H. (2001), A\&A, 365, 360

\bibitem[Thilker et al.(2004)]{thi04} Thilker D.A., Braun R., Walterbos R.A.M., Corbelli E., Lockman F.J., Murphy E., Maddalena R. (2004), ApJ, 601 39L

\bibitem[Thilker et al.(2007)]{thi07} Thilker D.A, Bianchi L., Meurer G., Gil de Paz A., Boissier S., Madore B.F., Boselli A., Ferguson A.M.N., Muńoz-Mateos J.C., Madsen G.J., Hameed S., Overzier R.A., Forster K., Friedman P.G., Martin D.C., Morrissey P., Neff S.G., Schiminovich D., Seibert M., Small T., Wyder T.K., Donas J., Heckman T.M., Lee Y.-W., Milliard B., Rich R.M., Szalay A.S., Welsh B.Y., Yi S.K. (2007), ApJS, 173, 538

\bibitem[Tinsley(1980)]{tinsley80} Tinsley B.M. (1980), {\it Fundamentals of Cosmic Physics}, Volume 5, pp. 287-388

\bibitem[Tinsley(1981)]{tinsley81} Tinsley B.M. (1981), ApJ, 250, 758

\bibitem[Tosi(1988)]{tosi88} Tosi M. (1988), A\&A 197, 33

\bibitem[Twarog(1980)]{twarog} Twarog B.A. (1980), ApJ, 242, 242

\bibitem[van der Hulst \& Sancisi(1988)]{hulst88} van der Hulst J.M.\ \& Sancisi R. (1988), AJ, 95, 1354

\bibitem[van der Hulst, van Albada \& Sancisi(2001)]{hulst01} van der Hulst J.M., van Albada T.S., Sancisi R. (2001), in: eds.\ J.E. Hibbard, M. Rupen \& J.H. van Gorkom, {\it Gas and Galaxy Evolution}, ASP Conf.\ Series, Vol.\ 240, p.\ 451

\bibitem[van der Hulst \& Sancisi(2005)]{hulst05} van der Hulst J.M.\ \& Sancisi R. (2005), in: ed.\ R.\ Braun, {\it Extra-planar gas}, ASP Conf.\ Series, Vol.\ 331, p.\ 139

\bibitem[van Gorkom \& Schiminovich(1997)]{vgork97} van Gorkom J.H.\ \& Schiminovich D. (1997), in: eds.\ M.\ Arnaboldi, G.S.\ Da Costa \& P.\ Saha, {\it The Nature of Elliptical Galaxies}, ASP Conf.\ Series, Vol.\ 116, p.\ 310

\bibitem[van Woerden et al.(2004)]{woerden04} van Woerden H., Wakker B.P., Schwarz U.J., de Boer K.S. (2004), {\it High-Velocity Clouds}, Astrophysics and Space Science Library, vol.\ 312, Kluwer Ac.Pub.(Dordrecht), eds.\ H.\ van Woerden, B.P.\ Wakker, U.J.\ Schwarz \& K.S.\ de Boer

\bibitem[van Woerden \& Wakker (2004)]{woerwak04} van Woerden H.\ \& Wakker B.P. (2004), in: eds.\ H.\ van Woerden, B.P.\ Wakker, U.J.\ Schwarz \& K.S.\ de Boer, {\it High-Velocity Clouds}, Astrophysics and Space Science Library, vol.\ 312, p.\ 195

\bibitem[Verheijen(2001)]{ver01} Verheijen M.A.W.\ \& Sancisi R. (2001), A\&A, 370, 765

\bibitem[Wakker \& van Woerden(1997)]{wak97} Wakker B.P\ \& van Woerden H. (1997), ARA\&A, 35, 217

\bibitem[Wakker et al.(2007a)]{wak07} Wakker B.P., York D.G., Howk C., Barentine J.C., Wilhelm R., Peletier R.F., van Woerden H., Beers T.C., Ivezic Z., Richter P., Schwarz U.J. (2007), ApJ, 670, 113L

\bibitem[Wakker et al.(2008)]{wak08} Wakker B.P., York D.G., Wilhelm R., Barentine J.C., Richter P., Beers T.C., Ivezic Z., Howk J.C. (2008), ApJ, in press, (astro-ph/0709.1926)

\bibitem[White \& Frenk(1991)]{whi91} White S.D.M.\ \& Frenk C.S. (1991), ApJ, 379, 52

\bibitem[Yun, Ho \& Lo(1994)]{yun94} Yun M.S., Ho P.T.P., Lo K.Y. (1994), Nature, 372, 530

\bibitem[Zaritsky \& Rix(1997)]{zar97} Zaritsky D.\ \& Rix H. (1997), ApJ, 477, 118 

\bibitem[Zwaan(2001)]{zwa01} Zwaan M.A., Briggs F.H., Sprayberry D. (2001), MNRAS, 327, 1249

\bibitem[Zwaan(2005a)]{zwa05a} Zwaan M.A., Meyer M.J., Staveley-Smith L., Webster R.L. (2005), MNRAS, 359, 30L
	
\bibitem[Zwaan(2005b)]{zwa05b} Zwaan M.A., van der Hulst J.M., Briggs F.H., Verheijen M.A.W., Ryan-Weber E.V. (2005), MNRAS, 364, 1467

\end{thebibliography}
\end{document}